\documentclass[journal,comsoc]{IEEEtran}
\usepackage[T1]{fontenc}

\ifCLASSINFOpdf
\else
   \usepackage[dvips]{graphicx}
   \graphicspath{{../eps/}}
   \DeclareGraphicsExtensions{.eps}
\fi
\usepackage{amsmath}
\usepackage[cmintegrals]{newtxmath}
\usepackage{cite}
\usepackage{amsfonts}
\usepackage{booktabs}
\usepackage{multirow}
\usepackage{subfigure}
\usepackage{tabularx}

\usepackage{algorithm}
\usepackage{algorithmic}
\usepackage{amsmath}

\usepackage{amssymb}
\usepackage{color}
\usepackage{array}
\usepackage{hyphenat}
\usepackage{url}
\usepackage{ntheorem}

\newtheorem{definition}{\textbf{Definition}}

\begin{document}
%
\title{A Distributed Virtual Network Function Placement Approach in Satellite Edge and Cloud Computing}
%
%
%

\author{Xiangqiang~Gao,
        Rongke~Liu,~\IEEEmembership{Senior~Member,~IEEE,}
        Aryan~Kaushik,~\IEEEmembership{Member,~IEEE,}
        and~Hangyu~Zhang
\thanks{X.~Gao, R.~Liu and H.~Zhang are with the School of Electronic and Information Engineering, Beihang University, Beijing 100191, China e-mail: (\{xggao, rongke\_liu, zhanghangyu\}@buaa.edu.cn).}
\thanks{A.~Kaushik is with the Department of Electronic and Electrical Engineering, University College London (UCL), London WC1E 7JE, United Kingdom e-mail: (a.kaushik@ucl.ac.uk).}}

\maketitle

\begin{abstract}
  Satellite edge computing has become a promising way to provide computing services for Internet of Things (IoT) devices in remote areas, which are out of the coverage of terrestrial networks, nevertheless, it is not suitable for large-scale IoT devices due to the resource limitation of satellites. Cloud computing can provide sufficient available resources for IoT devices but it does not meet the service requirements of delay sensitive users as high network latency. Collaborative edge and cloud computing is to facilitate flexible service provisioning for numerous IoT devices by incorporating the advantages of edge and cloud computing. In this paper, we investigate the virtual network function (VNF) placement problem in collaborative satellite edge and cloud computing to minimize the satellite network bandwidth usage and the service end-to-end delay. We formulate the VNF placement problem as an integer non-linear programming problem and propose a distributed VNF placement (D-VNFP) algorithm to address it, as the VNF placement problem is NP-hard. The experiments are conducted to evaluate the effectiveness of the proposed D-VNFP algorithm. The results show that the proposed D-VNFP algorithm outperforms the existing baseline algorithms of Greedy and Viterbi for solving the VNF placement problem in satellite edge and cloud computing.
\end{abstract}


%
\IEEEpeerreviewmaketitle

\section{Introduction}
%
%
%
%
\IEEEPARstart{T}{he} Internet of Things (IoT) has been widely applied in various fields, e.g., disaster monitoring, ocean transportation, target recognition and tracking, etc \cite{7289337}. Large amount of computation tasks produced by IoT devices need to be offloaded by terrestrial networks to remote servers for execution as IoT devices have the low computing capacity and battery power. However, terrestrial networks have not been set up in some remote areas due to high capital expenditures and harsh environments. Fortunately, low earth orbit (LEO) satellite networks, which have low communication delay and global coverage, can provide data collection, computing, and communication services for remote IoT devices without the coverage of terrestrial networks \cite{9222142}. In general, there are two computation offloading approaches in satellite-based IoT environment, where one is that computation tasks should be offloaded to ground cloud data centers by satellite networks for further processing \cite{8672604,9079470,9048610}, the other is that edge servers deployed on satellites can directly provide computing services for computation tasks \cite{8610431,9103927,9184934}.

For computation offloading in cloud computing, LEO satellite networks are responsible for collecting the data produced by remote IoT devices and transmitting them to ground cloud data centers via inter-satellite links (ISLs) \cite{9048610}. The computation tasks from remote IoT devices are processed by cloud servers and the processing results will be sent back by satellite networks to these remote IoT devices. Cloud computing can provide sufficient available resources for remote IoT devices on-demand by introducing software-defined network (SDN) and network function virtualization (NFV) \cite{6463372,9224163}. Nevertheless, cloud data centers are usually deployed far away from IoT devices. When computation tasks from IoT devices are offloaded to cloud servers by satellite networks, it will lead to high network latency as well as heavy network load, which can significantly degrade their processing performance \cite{8675467}.

Satellite edge computing has been considered as a new paradigm to provide computing services on the side of near IoT devices \cite{8610431}. Specifically, we can deploy edge servers on LEO satellites and provide satellite edge computing services for remote IoT devices without the coverage of terrestrial networks, which can significantly improve the real-time processing performance of computation tasks \cite{9103927}. As a result, satellite edge computing is suitable to provide edge computing services for delay sensitive devices in some remote areas. However, LEO satellites have the resource limitation of computing, storage, and bandwidth due to satellite physical constraints (e.g., satellite volume, energy, weight, etc.) \cite{9079470}. That is, satellite edge computing can not support edge computing services very well for large-scale IoT services. If the resource requirements of IoT devices exceed the resource capacities of satellites, the deployment failure of tasks will happen \cite{8823046}.

Considering that cloud computing has sufficient available resources and can provide flexible service provisioning for IoT devices while edge computing can support computing services for delay sensitive devices, collaborative edge and cloud computing has been investigated by combining the advantages of edge and cloud \cite{8089336,8664595}. The collaborative edge and cloud computing framework is composed of IoT devices, edge servers, and cloud centers, where computing resources can be provided for computation tasks in a hierarchical way \cite{8790769}. Delay sensitive services can be preferentially offloaded to edge servers, while other services can be offloaded to cloud data centers by transmission networks for further processing.

Note that the resource limitation of satellites is more serious compared with traditional edge computing. In order to improve the resource utilization of satellite networks in satellite edge and cloud computing, NFV combined with SDN has been a promising approach in orchestrating virtual network functions (VNFs) and managing network resources, where NFV can facilitate the decoupling of software and hardware, enable network functions to run on commercial servers, and abstract the available resources of satellite networks to be a resource pool to provide computing services for IoT devices on-demand. Specifically, a computation task can be divided into several interdependent network functions to deploy on different satellites, the data between adjacent network functions will be steered by inter-satellite links and we need to guarantee that the data flow can traverse all network functions in specific order. Thus, a computation task can be carried out by multiple satellites in a cooperative way. In addition, for centralized network resource management approaches, there exists high processing delay in satellite edge computing environment due to the property of satellite networks and the real-time computing service can be not satisfied for IoT devices. Furthermore, centralized network resource management controllers on the ground can be easily destroyed as disastrous events happen, such as earth quake, tsunami, and war. As a result, decentralized network resource management approaches are worth to investigating in satellite edge and cloud computing. However, to the best of our knowledge, no existing work focuses on the VNF placement problem by decentralized optimization algorithms in collaborative satellite edge and cloud computing \cite{8672604}.
\begin{figure}[tbp]
  \centering
  \includegraphics[width = \columnwidth]{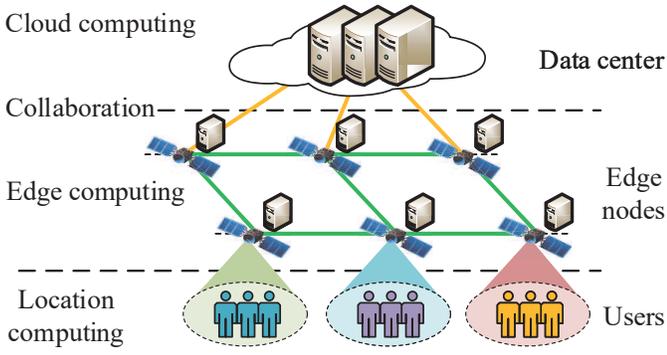}
  \caption{Hierarchical satellite edge and cloud computing architecture.}
  \label{Hierarchical satellite edge and cloud computing architecture}
\end{figure}

In our previous work \cite{Potentialgame_sat_IoT}, we proposed a potential game approach to address the VNF placement problem in satellite edge computing, where the satellite coverage for IoT devices is not considered as well as collaborative satellite edge and cloud computing. In this paper, we study the VNF placement problem in collaborative satellite edge and cloud computing, where the architecture consists of three-layer computing services including IoT devices, edge servers, and cloud servers, as shown in Fig.~\ref{Hierarchical satellite edge and cloud computing architecture}. IoT devices, which have low computing capacity and battery power, are distributed to offer environment observation and local computing in remote areas. Satellites can realize traffic routing and edge computing for computation tasks. Cloud servers can provide computing services for delay insensitive tasks. We consider a IoT device as a user and its task as a user request, which consists of several VNFs in order and is viewed as a service chaining \cite{7332796}.

For satellite edge computing, we can deploy the VNFs from a user request to multiple satellite edge servers to be executed in a collaborative way and the data between different VNFs will be routed by inter-satellite links. If user requests can not be deployed to edge servers as the resource limitation of satellites, then such deployment failed user requests will be offloaded by satellite networks to cloud servers for further processing. We assume that a IoT user can be covered by multiple satellites and thus there are various strategies related to user access and VNF placement, which have a significant influence on the performance in terms of network bandwidth usage and user end-to-end delay, so we consider the impact of different satellite coverage areas on the VNF placement solution. In addition, if a user request is not covered by any satellites or the resource requirements are not satisfied, then the user request will be considered to carry out locally.

In collaborative satellite edge and cloud computing, we formulate the VNF placement problem as an integer non-linear programming problem and our aim is to minimize the satellite network bandwidth usage and the user end-to-end delay. Considering that centralized optimization approaches have high processing delay for satellite edge and cloud computing, we propose a distributed virtual network function placement (D-VNFP) algorithm to address the VNF placement problem, which is viewed as NP-hard \cite{7469866,9180293}. For the proposed D-VNFP algorithm, each satellite is considered as an agent and can independently make the VNF deployment decision for each user request according to the current satellite network state and the resource requirements. These deployment strategies for user requests can be performed by competing available resources with each other in parallel. In this instance, some user requests may experience deployment failure as the available resources of satellites are used by other user requests. For such deployment failed user requests, we can update the satellite network resource state and then run the D-VNFP algorithm again to obtain new deployment strategies. When all user requests are deployed or any user request can no longer be deployed due to the resource limitation of satellites, the D-VNFP algorithm will terminate. In order to evaluate the performance of the proposed D-VNFP algorithm, the experiments are conducted in collaborative satellite edge and cloud computing environment. The main contributions are summarized as follows.
\begin{itemize}
  \item We study the VNF placement problem in collaborative satellite edge and cloud computing and build the hierarchical satellite edge and cloud computing framework.
  \item We formulate the VNF placement problem as an integer non-linear programming problem to minimize the network bandwidth usage and the user end-to-end delay. Then we propose a distributed virtual network function placement algorithm to tackle the problem.
  \item We evaluate the performance of the proposed D-VNFP algorithm in collaborative satellite edge and cloud computing environment. The results show that the proposed D-VNFP algorithm has better performance than the two existing baseline algorithms of Greedy and Viterbi.
\end{itemize}

The remainder of this paper is organized as follows. Section \ref{Related Work} briefly reviews related work concerning satellite-based IoT, satellite edge computing, and collaborative edge and cloud computing. Section \ref{System Model} introduces the system model in collaborative satellite edge and cloud computing. We formulate the VNF placement problem in Section \ref{Problem Formulation}. We propose a distributed virtual network function placement algorithm for addressing the VNF placement problem in Section \ref{Proposed Algorithm}. In Section \ref{Performance Evaluation}, we evaluate the performance of the proposed D-VNFP algorithm in static and dynamic environments. Finally, we provide the conclusion of this paper in Section \ref{Conclusion}.\par

\section{Related Work}\label{Related Work}
In this section, we firstly introduce satellite-based IoT applications. Then we present the related work about computation offloading in satellite edge computing. Finally, we review the literature related to collaborative edge and cloud computing.

\subsection{Satellite-based IoT}
Satellite networks can provide global seamless coverage and agile service provisioning for remote IoT devices, where satellite networks can provide access and transmission services for IoT devices which are lack of terrestrial networks \cite{8672604,9079470,9222142,8768353,9222141,8957684,8681409}. LEO satellites, which have low deployment cost and transmission delay, make a wide range of IoT services possible, such as remote sensing, earth observation, and ocean transportation \cite{9079470}. Considering that computation tasks of IoT devices need to be offloaded to cloud servers, the authors in \cite{9222142} proposed an energy efficient random access scheme for IoT devices to minimize energy consumption of user sides, the authors in \cite{8768353} proposed a random pattern multiplexing approach for random access to improve the performance of satellite-based IoT networks. The authors in \cite{8681409} presented the green data collection approach by LEO satellites in geo-distributed IoT networks. The existing work did not discuss satellite edge computing. In this paper, we study the problem of edge and cloud computing in satellite-based IoT networks.

\subsection{Satellite Edge Computing}
Satellite edge computing can provide flexible edge computing services for remote IoT devices by deploying edge servers on LEO satellites and introducing network virtualization \cite{8610431,9103927,9184934}. In \cite{9048610}, the authors discussed an architecture of satellite-terrestrial integrated edge computing networks, where they integrated mobile edge computing with the satellite network in order to implement satellite edge computing, and provided a task processing procedure in detail. The authors in \cite{9103927} jointly optimized the resource allocation of communication and computation for mobile edge computing to minimize the resource utilization and the end-to-end delay in maritime networks, and designed a deep reinforcement learning approach to address the problem. The authors in \cite{8675467} considered that computing servers can be deployed on edge nodes (e.g., LEO satellites and base stations) in software-defined satellite-terrestrial networks and used a deep Q-Learning approach to jointly optimize the resource allocation of network, cache, and computation. The authors in \cite{9184934} discussed computation offloading in satellite-based vehicle-to-vehicle communication systems, where they considered the optimization problem of offloading decision, computing and communication resource allocation to minimize the end-to-end delay cost and proposed a markov decision process approach to tackle the problem. For these existing work, a computation task is considered as a whole to allocate network resources. In this paper, we assume that a user request consists of several VNFs in order and is viewed as a service chaining, and the VNF placement problem is investigated to jointly optimize the network bandwidth usage and the user end-to-end delay.

\subsection{Collaborative Edge and Cloud Computing}
Collaborative edge and cloud computing can significantly improve the quality of experience for IoT users and enhance their processing performance \cite{8790769,8664595,9099432,9001206,8626532}. In \cite{8790769}, the authors proposed a three-tier edge-cloud collaborative residential energy management architecture for reducing network latency and energy cost, and presented a priority-based demand ratio scheduling strategy to deal with the energy management problem. The authors in \cite{8664595} discussed a joint resource allocation problem of communication and computation while minimizing the network latency in collaborative edge and cloud computing. The authors in \cite{9001206} proposed a dynamic deep-learning-based virtual edge node placement scheme in an edge and cloud environment, where the pay-as-you-go and spot instance model of cloud computing was applied to provide computing resources with minimum cost, and they used the long short-term memory in order to predict the information of user requests and resource prices. The authors in \cite{8626532} jointly optimized offloading decision, transmission power, and radio resource allocation on the uplink channel in multi-tier edge and cloud computing, and minimized the network computational cost and the user energy consumption. However, in this paper, we discuss the VNF placement problem in collaborative satellite edge and cloud computing, where we jointly consider the coverage of satellites, VNF deployment, and traffic routing in the system model, and formulate the resource allocation problem of computing, communication, and storage. We also propose a decentralized virtual network function placement algorithm to address the problem.

\section{System Model}\label{System Model}
In this section, we firstly provide the hierarchical architecture of collaborative satellite edge and cloud computing and the computation offloading model. Then we discuss the user request model and the VNF placement problem in collaborative satellite edge and cloud computing.

\subsection{Satellite Edge and Cloud Computing Architecture}
\begin{figure}[tbp]
  \centering
  \includegraphics[width = \columnwidth]{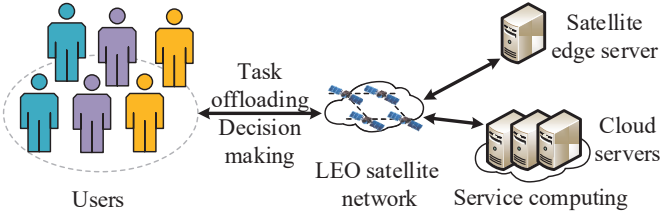}
  \caption{Different computation offloading approaches.}
  \label{Different computation offloading approaches}
\end{figure}
For the VNF placement problem in collaborative satellite edge and cloud computing, we consider a hierarchical computation offloading architecture, which includes IoT users, LEO satellites, and a ground cloud data center, as shown in Fig.~\ref{Hierarchical satellite edge and cloud computing architecture}. Local IoT users have low computing capacity and battery power, and are located in some remote areas without the coverage of terrestrial networks. The observation data from IoT users can be collected by LEO satellites and transmitted by inter-satellite links to satellite edge servers and a ground cloud data center for further processing. Edge servers on LEO satellites can provide computing services for delay sensitive user requests on the side of users. A ground cloud data center can offer sufficient resource provisioning for user requests on-demand but there exists high processing delay due to the property of satellite networks. Fig.~\ref{Different computation offloading approaches} describes the computation offloading strategy in collaborative satellite edge and cloud computing. Due to the disadvantages of local user computing and cloud computing, satellite edge computing has a high priority to deploy computation tasks. When the resource requirements of user requests are more than the resource limitation of satellites, we consider to offload user requests to the remote cloud data center for further processing while guaranteeing their service quality. However, some user requests will have deployment failure as they are not in the coverage of satellites or their resource requirements can not be satisfied by satellite edge and cloud, deployment failed user requests will be performed locally \cite{8823046}. Furthermore, centralized controllers are not used to orchestrate service functions and manage network resources in the proposed collaborative satellite edge and cloud architecture. To this end, we present a distributed resource allocation algorithm to manage the network resources and deploy VNFs for user requests. We assume that the network state and the user deployment strategy can be shared by a message synchronization mechanism.

We denote the LEO satellite network as a graph $G(V,E)$, where the parameters $V$ and $E$ represent the set of satellites and the set of inter-satellite links, respectively. The number of satellites in the set $V$ is $N$ and $v_n$ indicates the $n$-th LEO satellite. The set of resource types supported by the satellite $v_n$ is indicated as $R_n=\{CPU,memory\}$ due to that we just consider central processing unit (CPU) and memory for computing service in this paper. The $r$-th resource capacity for the satellite $v_n$ is expressed by $C_{n}^{r}$, where the parameter $r$ represents the supported resource type, e.g., CPU and memory. There are four inter-satellite links for a satellite, in which two links are from adjacent satellites in the same orbit and the other two links are from inter-orbit adjacent satellites. For the inter-satellite link $e, \forall e \in E$, we indicate the bandwidth capacity as $B_{e}$ and the transmission delay as $t_{e}$, respectively. Furthermore, we assume that there are sufficient resources for user requests in the cloud data center $v_{dc}$. We use $e_{n,dc}$ to indicate the link between the satellite $v_n$ and the cloud data center $v_{dc}$, $B_{n,dc}$ to indicate the bandwidth capacity, and $t_{n,dc}$ to indicate the transmission delay. We also assume that the cloud data center is in the coverage of at least one LEO satellite and can communicate with these satellites, which cover the cloud data center, by the satellite-ground link.

\subsection{User Requests}
The set of user requests is represented as $U$, where the number of user requests is $M$ and the $m$-th user request is indicated as $u_m$. We describe a user request as a service chaining, which is composed of multiple VNFs in particular order, the user request $u_m$ can be considered as a directed graph $G(F_m,H_m)$. We use $F_m$ to indicate the set of VNFs for the user request $u_m$, where the VNF $f_{m,i}$, $f_{m,i} \in F_m$, represents the $i$-th VNF of the user request $u_m$, the VNF $f_{m,1}=f_{m,s}$ represents the source access function, and the VNF $f_{m,\left | F_m\right |}=f_{m,d}$ represents the destination access function. For the user request $u_m$, we indicate the source $s_m$ as a sensor to collect the data and the destination $d_m$ as an actuator to interact with the environment by the processing results. We assume that the source $s_m$ and the destination $d_m$  may be different but known in advance. All the VNFs except the source access function and the destination access function have the requirements of CPU, memory, and the computing time. We denote the $r$-th resource requirements for the VNF $f_{m,i}$ as $c_{m,i}^{r}$ and the computing time as $t_{m,i}$, respectively. We use $H_m$ to indicate the set of edges for the user request $u_m$, where we indicate the edge between the adjacent VNFs $f_{m,i_1}$ and $f_{m,i_2}$ as $h_{m}^{i_1,i_2}$ and the bandwidth requirements of the edge $h_{m}^{i_1,i_2}$ as $b_{m}^{i_1,i_2}$. The maximum acceptable delay for the user request $u_m$ is denoted as $t_{m}^{max}$. Supposing that all IoT users are distributed in remote areas that are out of the coverage of terrestrial networks, only these IoT users in the coverage of satellites can offload their computation tasks to satellite edge and cloud servers. Each IoT user can communicate with these satellites which cover the IoT user, IoT users without the coverage of satellites will perform their tasks locally. We consider the number of user requests to come in a time slot as a Poisson distribution $P(\lambda_P)$ and the running time of each IoT user request as an Exponential distribution $E(\lambda_E)$.

\subsection{VNF Placement in Satellite Edge and Cloud Computing}
\begin{figure}[tbp]
  \centering
  \includegraphics[width = \columnwidth]{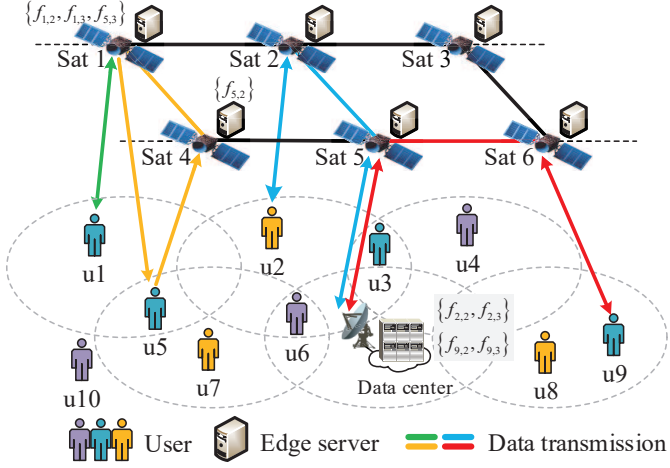}
  \caption{Example of deploying user requests in satellite edge and cloud.}
  \label{Example of deploying user requests in satellite edge and cloud}
\end{figure}
In this paper, we discuss the VNF placement problem in collaborative satellite edge and cloud computing, where our aim is to deploy $M$ user requests $U=\left \{ u_1,u_2,\cdots,u_M \right \}$ to $N$ satellite edge servers $V=\left \{ v_1,v_2,\cdots,v_N \right \}$ and a cloud data center $v_{dc}$ during a time slot while minimizing the network bandwidth usage and the user end-to-end delay. Similar to existing work on edge and cloud computing \cite{8823046,9179773,9091617}, we consider the VNF placement problem in a quasi-static scenario, where both the LEO satellite network topology and the set of user requests remain unchanged during the computing service period. The number and service requirements of user requests should vary over different computing service periods. All available satellite network resources are virtualized to provide service provisioning for user requests on-demand. Then we use a batch processing model to deploy these user requests to satellite edge and cloud servers during a time slot in dynamic environment.

For the VNF placement during a time slot, we firstly collect the new user requests to appear in the next time slot and update the current satellite network state as per the availability of the resources used by the completed user requests in the previous time slot. As IoT users have low computing capacity and battery power, the procedure of making the VNF placement decisions can be performed on LEO satellites to improve the real-time processing performance. To be more specific, we send the access information of each user request to one of the source neighbouring satellites and make the VNF placement decision for the user request based on the current satellite network state and the resource requirements, where the neighbouring satellite can be defined as follows.
\begin{definition}\label{definition1}
(\textbf{Neighbouring Satellite}) The user request $u_m$ includes a source and a destination. The source is in the coverage of LEO satellites $N(u_m,s)$, then we consider that these LEO satellites $N(u_m,s)$ are the source neighbouring satellites for the user request $u_m$. Similarly, the destination is in the coverage of LEO satellites $N(u_m,d)$, then we consider that these LEO satellites $N(u_m,d)$ are the destination neighbouring satellites for the user request $u_m$.
\end{definition}

The VNF placement strategy with minimum network bandwidth usage and end-to-end delay will be sent back to the user. These source neighbouring satellites can share the VNF placement strategy by a message synchronization mechanism to build the service chaining for the user request, where the VNF $f_{m,s}$ is deployed on one source neighbouring satellite as an access function for offloading the task to satellite edge and cloud, and the VNF $f_{m,d}$ is deployed on one destination neighbouring satellite as an access function for forwarding the processing results to the destination. However, different VNF placement strategies for a user request have a significant influence on the deployment cost in terms of network bandwidth usage and user end-to-end delay.

In order to better describe the VNF placement problem in collaborative satellite edge and cloud computing, we provide an example of deploying user requests in collaborative satellite edge and cloud computing, as shown in Fig.~\ref{Example of deploying user requests in satellite edge and cloud}. There are 6 LEO satellites, e.g., $v_1,v_2,\cdots,v_6$. Each LEO satellite, which hosts an edge server, can cover a particular ground area and provide computing services for user requests. These LEO satellites are connected with inter-satellite links and can steer the data flow from user requests with each other. There are also 10 user requests, e.g., $u_1,u_2,\cdots,u_{10}$, and a cloud data center $v_{dc}$ distributed in the ground area, where the data center $v_{dc}$ is in the coverage of the satellite $v_5$. Considering that a user request can just communicate with its neighbouring satellites, the VNFs $f_{1,s},f_{1,2},f_{1,3}$, and $f_{1,d}$ for the user request $u_1$, which is in the coverage of the satellite $v_1$, will be deployed to the satellite $v_1$. The user request $u_5$ is in the coverage of the satellites $v_1$ and $v_4$, and can communicate with the two neighbouring satellites, respectively. The user request $u_5$ can communicate with the satellite $v_4$ to deploy the VNF $f_{5,2}$ to the satellite $v_4$ and the VNF $f_{5,3}$ to the satellite $v_1$, where we deploy the VNFs $f_{5,s}$ and $f_{5,d}$ to the satellites $v_4$ and $v_1$, respectively, the processing results will be sent back to the user request $u_5$ by the satellite $v_1$. When the available resources of the satellite network are less than the resource requirements of user requests, these user requests will be offloaded to the cloud data center $v_{dc}$ by the satellite network. For example, the user request $u_9$ can be offloaded to the cloud data center $v_{dc}$ by the satellites $v_6$ and $v_5$, and the user request $u_2$ can be offloaded to the cloud data center $v_{dc}$ by the satellites $v_2$ and $v_5$. The VNFs $f_{2,2}$, $f_{2,3}$, $f_{9,2}$, and $f_{9,3}$ for the two user requests $u_2$ and $u_9$ can be executed in the cloud data center, where the VNFs $f_{2,s}$ and $f_{2,d}$ are deployed to the satellite $v_2$, and the VNFs $f_{9,s}$ and $f_{9,d}$ are deployed to the satellite $v_6$. However, the user request $u_{10}$ will be carried out locally as it is out of the coverage of any satellites.

\section{Problem Formulation}\label{Problem Formulation}
\begin{table}[tbp]
  \caption{List of Symbols.}
  \label{List of Symbols}
  \centering
  \resizebox{\columnwidth}{!}{
  \begin{tabular}{cl}
  \hline
  \multicolumn{2}{c}{\bfseries Satellite Network}\\
  \hline
  $G(V,E)$ & The satellite network with satellites $V$ and links $E$.\\
  $R_n$ & The set of resources offered by satellite $v_n$, $v_n \in V$.\\
  $C_{n}^{r}$ & The $r$-th resource capacity of satellite $v_n$, $r \in R_n$.\\
  $B_{e},t_{e}$ & Bandwidth and delay of link $e$, $e \in E$.\\
  $v_{dc}$ & The cloud data center.\\
  $B_{n,dc}$ & Bandwidth capacity of the link between $v_n$ and $v_{dc}$.\\
  $P_{n_{1}}^{n_{2}}$ & Set of the $d$ shortest paths between $v_{n_{1}}$ and $v_{n_{2}}$.\\
  \hline
  \multicolumn{2}{c}{\bfseries User Requests}\\
  \hline
  $U$ & The set of $M$ user requests.\\
  $F_m,H_m$ & VNFs $F_m$ and edges $H_m$ of user request $u_m$.\\
  $f_{m,i}$ & The $i$-th VNF of user request $u_m$.\\
  $s_m,d_m$ & Source and destination of user request $u_m$.\\
  $c_{m,i}^{r}$ & The $r$-th resource requirements of $f_{m,i}$.\\
  $t_{m,i}$ & The computing time of $f_{m,i}$.\\
  $h_{m}^{i_{1},i_{2}}$ & Edge between $f_{m,i_{1}}$ and $f_{m,i_{2}}$.\\
  $b_{m}^{i_{1},i_{2}}$ & The bandwidth requirements of edge $h_{m}^{i_{1},i_{2}}$. \\
  $t_{m}^{max}$ & Maximum acceptable delay of user request $u_m$. \\
  \hline
  \multicolumn{2}{c}{\bfseries Binary Variables}\\
  \hline
  $x_{m}$ & $x_{m} = 1$ if $u_m$ is deployed to satellite edge and cloud.\\
  $y_{m}$ & $y_{m} = 1$ if edge servers provide service for $u_m$.\\
  $z_{m,i}^{n}$ & $z_{m,i}^{n} = 1$ if $f_{m,i}$ is deployed to satellite $v_n$.\\
  $w_{m,p}^{i_{1},i_{2}}$ & $w_{m,p}^{i_{1},i_{2}}=1$ if path $p$ is used by $h_{m}^{i_{1},i_{2}}$.\\
  $w_{e}^{p}$ & $w_{e}^{p} = 1$ if link $e$ is used by path $p$.\\
  \hline
  \multicolumn{2}{c}{\bfseries Objective Variables}\\
  \hline
  $C_{bw}$ & Average satellite network bandwidth cost.\\
  $C_{user}$ & Average user end-to-end delay cost.\\
  \hline
  \end{tabular}
  }
\end{table}
In this section, we formulate the VNF placement problem to jointly minimize the network bandwidth usage and the user end-to-end delay in collaborative satellite edge and cloud computing. User requests distributed in some remote areas are deployed to satellite edge servers and cloud servers while minimizing the satellite network bandwidth usage and the user end-to-end delay, and guaranteeing the service quality of user requests. The main symbols for the proposed VNF placement problem are summarized in Table \ref{List of Symbols}.

We use a binary variable $x_m=\{0,1\}$ to indicate whether the user request $u_m$ is deployed to satellite edge and cloud servers. The variable $x_m=1$ if the user request $u_m$ is deployed to satellite edge and cloud servers, otherwise $x_m=0$.

We assume that the user request $u_m$ is successfully deployed to satellite edge and cloud servers, then we use another binary variable $y_m=\{0,1\}$ to indicate whether satellite edge servers provide computing service for the user request $u_m$. The variable $y_m=1$ if satellite edge servers provide computing service for the user request $u_m$, otherwise $y_m=0$.

If the user request $u_m$ is deployed to edge servers, we use a binary variable $z_{m,i}^{n}=\{0,1\}$ to indicate whether the VNF $f_{m,i}$ is deployed to the satellite $v_n$. The variable $z_{m,i}^{n}=1$ if the VNF $f_{m,i}$ is deployed to the satellite $v_n$, otherwise $z_{m,i}^{n}=0$.

We assume that the $d$ shortest paths between any two satellites $v_{n_1}$ and $v_{n_2}$ are known in advance and the path set can be indicated as $P_{n_1}^{n_2}$. We use a binary variable $w_{m,p}^{i_1,i_2}=\{0,1\}$ to indicate whether the path $p$ is used by the edge $h_{m}^{i_1,i_2}$. The variable $w_{m,p}^{i_1,i_2}=1$ if the path $p$ is used by the edge $h_{m}^{i_1,i_2}$, otherwise $w_{m,p}^{i_1,i_2}=0$.

For the path $p$, it consists of multiple inter-satellite links. We use a binary variable $w_e^p=\{0,1\}$ to indicate whether the link $e$ is used by the path $p$. The variable $w_e^p=1$ if the link $e$ is used by the path $p$, otherwise $w_e^p=0$.

Furthermore, the source of a user request can just only communicate with its neighbouring satellites. We assume that the source of the user request $u_m$ is in the coverage of the satellite $v_n$, that is
\begin{equation}\label{equation1}
s_m \in cov(v_n), \forall u_m \in U, \forall v_n \in V,
\end{equation}
where the parameter $cov(v_n)$ indicates the coverage of the satellite $v_n$. Then the set $N(u_m,s)$ of neighbouring satellites for the source of the user request $u_m$ can be described as
\begin{equation}\label{equation2}
N(u_m,s) = \left\{ {\left. {{v_n}} \right|{s_m} \in cov(v_n)} \right\},\forall {v_n} \in V.
\end{equation}
In the same way, the set $N(u_m,d)$ of neighbouring satellites for the destination of the user request $u_m$ can be described as
\begin{equation}\label{equation3}
N(u_m,d) = \left\{ {\left. {{v_n}} \right|{d_m} \in cov(v_n)} \right\},\forall {v_n} \in V.
\end{equation}
We also denote the neighbouring satellites for the cloud data center $v_{dc}$ as \begin{equation}\label{equation4}
N(dc) = \left\{ {\left. {{v_n}} \right|{v_{dc}} \in cov(v_n)} \right\},\forall {v_n} \in V.
\end{equation}
When user requests are deployed to satellite edge servers and cloud servers, these neighbouring satellites for user requests and the data center will be considered as access nodes to offload computation tasks and receive the processing results, where the VNFs $f_{m,s}$ and $f_{m,d}$ for the user request $u_m$ will be deployed the source and destination neighbouring satellites to rout the traffic flow, respectively. So we need to specify which one neighbouring satellite is used to be an access node in satellite edge and cloud computing.

For satellite edge and cloud computing, a binary variable $q_{m,s}^{n}=\{0,1\}$ is used to indicate whether the source neighbouring satellite $v_n$ is an access node for the user request $u_m$ to offload the task, the variable $q_{m,s}^{n}=1$ if the neighbouring satellite $v_n$ is an access node for the user request $u_m$ to offload the task, otherwise $q_{m,1}^{n}=0$. Similarly, another binary variable $q_{m,d}^{n}=\{0,1\}$ is used to indicate whether the destination neighbouring satellite $v_n$ is an access node for the user request $u_m$ to obtain the processing results, the variable $q_{m,d}^{n}=1$ if the neighbouring satellite $v_n$ is an access node for the user request $u_m$ to obtain the processing results, otherwise $q_{m,d}^{n}=0$.

For cloud computing, the satellite network is just considered as a communication network and can not provide edge computing services for user requests, all computation tasks will be offloaded to the cloud data center for further processing. We use a binary variable $q_{m,s}^{n,dc}=\{0,1\}$ to indicate whether the neighbouring satellite $v_n$ of the data center is an access node for the user request $u_m$ to offload the task to the data center, the variable $q_{m,s}^{n,dc}=1$ if the neighbouring satellite $v_n$ of the data center is an access node for the user request $u_m$ to offload the task to the data center, otherwise $q_{m,s}^{n,dc}=0$. We use another binary variable $q_{m,d}^{n,dc}=\{0,1\}$ to indicate whether the neighbouring satellite $v_n$ of the data center is an access node for the user request $u_m$ to send back the processing results, the variable $q_{m,d}^{n,dc}=1$ if the neighbouring satellite $v_n$ of the data center is an access node for the user request $u_m$ to send back the processing results, otherwise $q_{m,d}^{n,dc}=0$.

When we deploy a user request to satellite edge servers and cloud servers, there are multiple physical constraints to be considered, such as network resource capacity, user service quality, and deployment implementation.

If a user request is offloaded to satellite edge servers and cloud servers, we need to indicate one source neighbouring satellite to offload the task and one destination neighbouring satellite to receive the processing results, respectively. Therefore, it is guaranteed that only one source and destination neighbouring satellite for the user request $u_m$ can be used. For $\forall {u_m} \in U,\forall {v_{n_1}} \in N(m,s),\forall {v_{n_2}} \in N(m,d)$, the access constraint of neighbouring satellites can be represented as
\begin{equation}\label{equation5}
{x_m} \cdot (\left| {1 - \sum\limits_{{v_{{n_1}}}} {q_{m,s}^{n_1}} } \right| + \left| {1 - \sum\limits_{{v_{{n_2}}}} {q_{m,{d}}^{n_2}} } \right|) = 0.
\end{equation}
When the user request $u_m$ is deployed to satellite edge servers, we consider that each VNF $f_{m,i}$ can be deployed to only one edge server. For $\forall {u_m} \in U$, the VNF placement constraint can be indicated as
\begin{equation}\label{equation6}
x_m \cdot y_m \cdot (1 - \sum\limits_{{v_n}} {z_{m,i}^n} ) = 0,\forall {f_{m,i}} \in {F_m},\forall {v_n} \in V.
\end{equation}
For the user request $u_m$ offloaded to satellite edge servers, when the two adjacent VNFs $f_{m,i_1}$ and $f_{m,i_2}$ are deployed to the edge servers on the satellites $v_{n_1}$ and $v_{n_2}$, respectively, it needs to be guaranteed that one of the $d$ shortest paths between the satellites $v_{n_1}$ and $v_{n_2}$ can be used to route the traffic flow from the VNF $f_{m,i_1}$ to the VNF $f_{m,i_2}$. The path selection constraint can be expressed as
\begin{equation}\label{equation7}
x_m \cdot y_m \cdot ( z_{m,{i_1}}^{{n_1}} \cdot z_{m,{i_2}}^{{n_2}} - \sum\limits_p {w_{m,p}^{{i_1},{i_2}}} ) = 0,\forall u_m \in U,\forall p \in P_{{n_1}}^{{n_2}}.
\end{equation}
As satellite edge servers have the limited resource capacity in terms of CPU and memory, the resource requirements of user requests for each satellite should be less than the available resources. The $r$-th used resources of the satellite $v_n$ is indicated as $c_{n,used}^{r}$. For $\forall {v_n} \in V,\forall r \in {R_n}$, we can indicate the satellite resource constraint as
\begin{equation}\label{equation8}
c_{n,used}^{r}\! +\!\! \sum\limits_{{u_m}} y_m \!{\sum\limits_{{f_{m,i}}} {z_{m,i}^n \!\cdot\! c_{m,i}^r} }  \!\!\le\!\! C_n^r, \forall {u_m} \!\in\! U, \forall {f_{m,i}} \!\in\! {F_m}.
\end{equation}

Furthermore, we also need to guarantee that the bandwidth requirements of user requests should not exceed the available bandwidth resources. When user requests are deployed to satellite edge servers, where $\forall u_m \in U$, $\forall h_{m}^{i_1,i_2} \in H_{m},\forall v_{n_1},v_{n_2} \in V$, and $\forall p \in P_{n_1}^{n_2}$, the bandwidth requirements $b_{e,1}$ of the link $e$, $\forall e \in E$, can be represented as
\begin{equation}\label{equation9}
b_{e,1} \!= \!\!\sum\limits_{{u_m}} {\!{x_m \!\cdot \!y_m}\!\!\sum\limits_{h_m^{{i_1},{i_2}}} {\sum\limits_{{v_{{n_1}}},{v_{{n_2}}}} {\!\!\!z_{m,{i_1}}^{{n_1}} \!\!\cdot\!  z_{m,{i_2}}^{{n_2}}\!\sum\limits_p \!{w_{m,p}^{{i_1},{i_2}} \!\cdot\! w_e^p \!\!\cdot\! b_m^{{i_1},{i_2}}} } } }.
\end{equation}
If the available resources of satellite edge servers are less than the resource requirements of user requests, we will consider to offload these unassigned user requests to the cloud data center for further processing while guaranteeing their service quality. To simplify the proposed VNF placement problem, we assume that all the VNFs of a user request except the source access function and the destination access function will be deployed to the cloud data center if a user request needs to be offloaded to the cloud data center. Specifically, the satellite network is responsible for routing the data flow from the source to the cloud data center and sending the processing results back from the cloud data center to the destination. When computation tasks of user requests are offloaded to the cloud data center, the bandwidth requirements for $\forall u_m \in U, \forall p \in P_{n_1}^{n_2}$, and $\forall {v_{n_1}} \in N(m,s),\forall {v_{n_2}} \in N(dc)$ can be indicated as
\begin{equation}\label{equation10}
b_{e,2}^1 = \sum\limits_{{v_{{n_1}}},{v_{{n_2}}}}{q_{m,s}^{{n_1}} \!\cdot\! q_{m,s}^{{n_2},dc}\sum\limits_p \!{w_{m,p}^{s,2}} \!\cdot\! w_e^p \!\cdot\! b_m^{s,{2}}}.
\end{equation}
When the processing results of user requests are sent back from the cloud data center to their destinations, the bandwidth requirements for $\forall u_m \in U,\forall p \in P_{n_1}^{n_2}$, and $\forall {v_{n_1}} \in N(dc),\forall {v_{n_2}} \in N(m,d)$ can be indicated as
\begin{equation}\label{equation11}
b_{e,2}^2 = \sum\limits_{{v_{{n_1}}},{v_{{n_2}}}}{q_{m,d}^{{n_1},dc} \!\cdot\! q_{m,d}^{{n_2}}\sum\limits_p \!{w_{m,p}^{\left|F_m\right|-1,d} \!\cdot\! w_e^p \!\cdot\! b_m^{\left|F_m\right|-1,d}}}.
\end{equation}
Then the bandwidth requirements for user requests, which are deployed to cloud servers, can be indicated as
\begin{equation}\label{equation12}
b_{e,2} = \!\!\sum\limits_{{u_m}} \!{{x_m} \!\cdot\! (1 \!-\! {y_m}) \!\cdot\!(b_{e,2}^1+b_{e,2}^2) }, \forall u_m \in U.
\end{equation}
We denote the used bandwidth resources of the link $e$ as $b_{e,used}$, then the bandwidth resource constraint of the link $e$ can be represented as
\begin{equation}\label{equation13}
b_{e,used} + b_{e,1} + b_{e,2} \le B_e, \forall e \in E.
\end{equation}
When the user request $u_m$ is deployed to cloud servers, we need to guarantee that there exists available neighbouring satellites for the cloud data center to offload the task and receive the processing results. The neighbouring satellite constraint for $\forall u_m \in U, \forall v_{n_1},v_{n_2} \in N(dc)$ can be described as
\begin{equation}\label{equation14}
{x_m} \cdot (1 - {y_m}) \cdot (\left| {1 - \sum\limits_{{v_{{n_1}}}} {q_{m,s}^{{n_1},dc}} } \right| + \left| {1 - \sum\limits_{{v_{{n_2}}}} {q_{m,d}^{{n_2},dc}} } \right|) = 0.
\end{equation}
We also ensure that there exists an available path between the user request $u_m$ and the cloud data center $v_{dc}$ by the satellite network in cloud computing, where the path select constraint for $\forall u_m \in U, \forall p \in P_{n_1}^{n_2}$ can be described as
\begin{equation}\label{equation15}
\left\{ {\begin{array}{*{20}{c}}
{{x_m} \cdot (1 - {y_m}) \cdot (q_{m,s}^{{n_1}} \cdot q_{m,d}^{{n_2},dc}{\rm{ - }}\sum\limits_p  w_{m,p}^{s,2}{\rm{) = }}0},\\
{{x_m} \cdot (1 - {y_m}) \cdot (q_{m,d}^{{n_1},dc} \cdot q_{m,d}^{{n_2}}{\rm{ - }}\sum\limits_p  w_{m,p}^{\left| {{F_m}} \right| - 1,d}{\rm{) = }}0}.
\end{array}} \right.
\end{equation}
Furthermore, the bandwidth requirements between the cloud data center $v_{dc}$ and the neighbouring satellite $v_n$, $v_n \in N(dc)$, should be less than the available bandwidth resources. We denote the used bandwidth resources of the link as $b_{n,dc}^{used}$, the bandwidth resource constraint for the satellite-ground link $e_{n,dc},v_n \in N(dc)$ can be indicated as
\begin{equation}\label{equation16}
b_{n,dc}^{used} + b_{n,dc}^{req} \le {B_{n,dc}},
\end{equation}
where $b_{n,dc}^{req}$ denotes the bandwidth requirements of user requests and can be indicated for $\forall u_m \in U$ as
\begin{equation}\label{equation17}
b_{n,dc}^{req} \!\!= \!\!\sum\limits_{{u_m}} {{x_m} \cdot (1 - {y_m})} \cdot (b_m^{s,2} \cdot q_{m,s}^{n,dc} + b_m^{\left|F_m\right|-1,d} \cdot q_{m,d}^{n,dc}).
\end{equation}

When we deploy user requests to satellite edge servers and the cloud data center, we also need to ensure that the processing delay of a user request, which is the sum of the computing delay and the transmission delay, can not exceed the maximum acceptable delay. The computing delay $t_m^{cmp}$ of the user request $u_m$ can be described as
\begin{equation}\label{equation18}
t_m^{cmp} = \sum\limits_{{f_{m,i}}} {{t_{m,i}}} ,{\forall u_m \in U, \forall f_{m,i}} \in {F_m}.
\end{equation}
We denote the transmission delay from the source to the source neighbouring satellite $v_n$ as $t_{s,n}$ and the transmission delay from the destination neighbouring satellite $v_n$ to the destination as $t_{n,d}$, respectively. When the user request $u_m$ is deployed to satellite edge servers, for $\forall v_{n_1} \in N(u_m,s), \forall v_{n_2} \in N(u_m,d)$, the transmission delay $t_{m,1}^{tr}$ can be indicated as
\begin{equation}\label{equation19}
t_{m,1}^{tr} = \sum\limits_{{v_{n_1}}} {q_{m,s}^{n_1} \cdot {t_{s,n_1}}}  + t_{m,1}^{sat}  + \sum\limits_{{v_{n_2}}} {q_{m,d}^{n_2} \cdot {t_{n_2,d}}},
\end{equation}
where the transmission delay of inter-satellite links for the user request $u_m$ is denoted as $t_{m,1}^{sat}$. For $\forall h_{m}^{i_1,i_2} \in H_m$, $\forall p \in P_{n_1}^{n_2}$, $\forall v_{n_1},v_{n_2} \in V$, and $\forall e \in p$, $t_{m,1}^{sat}$ can be described as
\begin{equation}\label{equation20}
t_{m,1}^{sat} = \!\!\sum\limits_{h_m^{{i_1},{i_2}}} {\sum\limits_{{v_{{n_1}}},{v_{{n_2}}}} \!\!\!{ z_{m,{i_1}}^{{n_1}} \!\!\cdot\!\! z_{m,{i_2}}^{{n_2}}\sum\limits_p {\sum\limits_e {w_{m,p}^{{i_1},{i_2}} \cdot {t_e}} } } }.
\end{equation}
To simplify the proposed VNF placement problem, we just consider the computing delay of the VNFs for user requests and neglect the processing delay of switches and links in the cloud data center. We assume that the transmission delay between the cloud data center and the neighbouring satellite $v_n$ for computation offloading is $t_{n,dc}$ and the transmission delay between the cloud data center and the neighbouring satellite $v_n$ for receiving the processing results is $t_{dc,n}$. When the user request $u_m$ is deployed to the cloud data center $v_{dc}$, we indicate the transmission delay $t_{m,2}^{tr,1}$ from the source to the cloud data center and the transmission delay $t_{m,2}^{tr,2}$ from the cloud data center to the destination as
\begin{equation}\label{equation21}
\left\{ {\begin{array}{*{20}{c}}
{t_{m,2}^{tr,1} = \sum\limits_{{v_{{n_1}}},{v_{{n_2}}}} {q_{m,s}^{{n_1}} \cdot q_{m,s}^{{n_2},dc}({t_{s,n_1}} + {t_{{n_2},dc}}) + t_{m,2}^{sat,1},} }\\
{t_{m,2}^{tr,2} = \sum\limits_{{v_{{n_1}}},{v_{{n_2}}}} {q_{m,d}^{{n_1},dc} \cdot q_{m,d}^{{n_2}}({t_{dc,n_1}} + {t_{{n_2},d}}) + t_{m,2}^{sat,2},} }
\end{array}} \right.
\end{equation}
where $t_{m,2}^{sat,1}$ and $t_{m,2}^{sat,2}$ for $\forall p \in P_{n_1}^{n_2}, \forall e \in p$ are indicated as
\begin{equation}\label{equation22}
\left\{ {\begin{array}{*{20}{c}}
{t_{m,2}^{sat,1}{\rm{ = }}\sum\limits_{{v_{{n_1}}},{v_{{n_2}}}} {q_{m,s}^{{n_1}} \cdot q_{m,s}^{{n_2},dc}\sum\limits_p  \sum\limits_e {w_{m,p}^{s,2} \cdot {t_e},} } }\\
{t_{m,2}^{sat,2}{\rm{ = }}\sum\limits_{{v_{{n_1}}},{v_{{n_2}}}} {q_{m,d}^{{n_1},dc} \cdot q_{m,d}^{{n_2}}\sum\limits_p  \sum\limits_e {w_{m,p}^{\left| {{F_m}} \right| - 1,d} \cdot {t_e},} } }
\end{array}} \right.
\end{equation}
then the transmission delay $t_{m,2}^{tr}$ for the user request $u_m$, which is deployed to cloud servers, can be indicated as
\begin{equation}\label{equation23}
t_{m,2}^{tr} =t_{m,2}^{tr,1} + t_{m,2}^{tr,2}.
\end{equation}
The total processing delay constraint of the user request $u_m$ can be expressed as
\begin{equation}\label{equation24}
t_{m}^{delay}=t_{m}^{cmp} + y_m \cdot t_{m,1}^{tr} + (1-y_m) \cdot t_{m,2}^{tr} \le t_{m}^{max}.
\end{equation}

In this paper, our optimization goal is to jointly minimize the average satellite network bandwidth usage and the average user end-to-end delay. According to the above discussion, we indicate the average satellite network bandwidth cost $C_{bw}$ as
\begin{equation}\label{equation25}
{C_{bw}} = \frac{1}{{\left| E \right|}}\sum\limits_e {(b_{e,used} + b_{e,1} + b_{e,2})}, \forall e \in E.
\end{equation}
We indicate the average user end-to-end delay cost $C_{user}$ as
\begin{equation}\label{equation26}
{C_{user}} = \frac{1}{\sum\limits_{u_m} x_m} \sum\limits_{u_m} {x_m \cdot t_{m}^{delay}},\forall {u_m} \in U.
\end{equation}
For the VNF placement optimization problem in collaborative satellite edge and cloud, we formulate two optimization sub-problems of the satellite network bandwidth cost in equation \eqref{equation25} and the user end-to-end delay cost in equation \eqref{equation26} within the physical network resource and service quality constraints in equations \eqref{equation1}-\eqref{equation24}. We can indicate the satellite network bandwidth cost optimization sub-problem as
\begin{equation}\label{equation27}
\begin{aligned}
\text{min}\quad & C_{bw} \\
s.t.\quad &  \eqref{equation1}-\eqref{equation24}.
\end{aligned}
\end{equation}
We can also indicate the user end-to-end delay cost optimization sub-problem as
\begin{equation}\label{equation28}
\begin{aligned}
\text{min}\quad & C_{user} \\
s.t.\quad &  \eqref{equation1}-\eqref{equation24}.
\end{aligned}
\end{equation}
In order to address the VNF placement problem, which consists of two sub-optimization problems in equations \eqref{equation27} and \eqref{equation28}, we put the user end-to-end optimization problem as the primary optimization objective and the network bandwidth optimization problem as the secondary optimization objective. When we find an optimal solution for the VNF placement problem, we need to firstly ensure the user end-to-end delay minimum and then optimize the network bandwidth cost.

\section{Proposed Algorithm}\label{Proposed Algorithm}
As shown in existing related work \cite{8187671,7332796,7469866,9180293}, the VNF placement problem is a NP-hard problem, where the classical NP-hard problems, e.g., Quadratic Assignment Problem (QAP) \cite{Rainer1984Quadratic} and Capacitated Plant Location Problem with Single Source Constraints (CPLPSS) \cite{1993A}, can be reduced to the VNF placement problem. Then the optimal solution for the VNF placement problem can only be obtained in small scale problems, but the time for finding the optimal VNF placement solution is unacceptable in large scale problems. Therefore, heuristic algorithms, e.g., DPVC \cite{8187671}, Greedy and Simulation Annealing \cite{7332796}, Viterbi \cite{7469866}, and Genetic Algorithm \cite{7885521}, have been widely used for the VNF placement problem to find an approximate solution in a low computational time.

In this section, we propose a distributed VNF placement (D-VNFP) heuristic algorithm to address the VNF placement problem of minimizing the satellite network bandwidth usage and the user end-to-end delay, as the VNF placement problem is NP-hard. For centralized VNF placement algorithms, satellite network resource state and user requests can be collected to a centralized controller for making the VNF placement decisions and then the strategy results will be sent back by satellite networks. For this reason, there is a high processing delay for making the VNF placement decisions in centralized VNF placement algorithms, especially for satellite networks. As a contrast, the proposed D-VNFP algorithm can make the VNF placement decisions for user requests on LEO satellites in a distributed manner and better provide real-time computing services for delay sensitive user requests. Then we analyze the computation complexity of the proposed D-VNFP algorithm.

\subsection{Distributed VNF Placement Algorithm}
For the proposed D-VNFP algorithm, we consider that the satellite $v_n$ hosts the agent $a_n$, which has the characteristics of environment-aware, autonomy, and social behavior, neighbouring satellites of user requests and the cloud data center serve as their neighbouring agents. The agents can share the network state by a message synchronization mechanism, make their own VNF placement decisions in a self-interested way, and perform user requests in a collaborative way \cite{zheng2015multi}. To better describe the proposed D-VNFP algorithm, we provide the following characteristics of the agents \cite{9224163}.
\begin{itemize}
  \item \emph{Environment-aware:} Each agent can obtain the current satellite network state including network topology and available network resources by a message synchronization mechanism. When user requests are to arrive in the next time slot, their neighbouring agents can also obtain the information concerning new user requests. Besides, the agents can acquire the VNF placement strategies of user requests by communicating with each other.
  \item \emph{Autonomy:} Each agent can make the own VNF placement decisions independently for user requests according to the current satellite network resource state. Furthermore, each agent can autonomously realize its resource management, data forwarding, and service function running depending on the VNF placement strategy.
  \item \emph{Social behavior:} Each agent can interact with other agents to obtain the satellite network state and the VNF placement strategy. When the neighbouring agent makes the VNF placement decision for a user request, its multiple neighbouring agents can share the VNF placement strategy. In addition, when a user request is deployed to multiple agents, these related agents can perform the user request in a collaborative way.
\end{itemize}
\begin{figure}[tbp]
  \centering
  \includegraphics[width = \columnwidth]{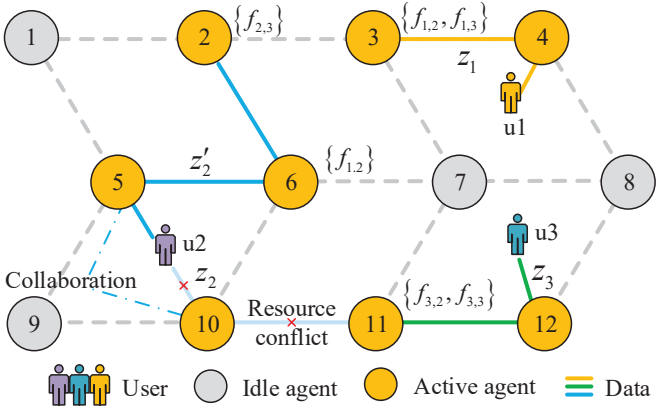}
  \caption{Procedure of the proposed D-VNFP algorithm.}
  \label{Procedure of the proposed D-VNFP algorithm}
\end{figure}

The procedure of the proposed D-VNFP algorithm includes two parts of strategy decision and service deployment. For the strategy decision, each agent firstly updates the local information about the current satellite network state by a message synchronization mechanism. When a new user request is to arrive, the access information will be randomly sent to one of the source neighbouring agents. The source neighbouring agent can make the VNF placement decision by the proposed D-VNFP algorithm, which jointly minimizes the network bandwidth usage and the user end-to-end delay. When a user request has multiple neighbouring agents, these neighbouring agents can share the VNF placement strategy with each other. It should be noted that the neighbouring agent, which makes the VNF placement decision, may be not the agent to steer the data produced by the user request.

For the service deployment, a user request usually needs multiple agents to execute in a collaborative way, that is, the VNF placement strategy for a user request may include multiple agents. Therefore, the neighbouring agent with the optimal VNF placement strategy needs to share the VNF placement strategy with other related agents. Based on the VNF placement strategy, we can place the VNFs to the corresponding agents, respectively, and build the service chaining. Each agent is responsible for running service functions and forwarding data independently, all related agents can collaborate to accomplish the user request. When a user request is end, we will break the service chaining and release the used resources, e.g., network bandwidth, CPU, and memory, to be available for new user requests in the next time slot.

Furthermore, the VNF placement decisions for all user requests are made on different agents concurrently and thus the agents does not know the VNF placement strategies with each other. As a result, there are potential resource conflicts, which will cause the resource requirements of user requests to exceed the available resources of the satellite network, for these VNF placement strategies of user requests. In order to address the potential resource conflict problem, we use a resource competition mechanism based on a first-come-first-serve procedure in the proposed D-VNFP algorithm, where each agent can give priority to providing the available resources for the VNF placement strategies that come first. When the required resources of user requests are greater than the agent available resources, then their neighbouring agents will update the local network resource state and perform the proposed D-VNFP algorithm again to find new approximate VNF placement strategies.

Fig.~\ref{Procedure of the proposed D-VNFP algorithm} illustrates an example of the proposed D-VNFP algorithm. There are $12$ satellite agents, e.g., $a_1,a_2,\cdots,a_{12}$, and each agent can directly communicate with four adjacent agents in the same orbital plane and adjacent one. Three user requests $u_1$, $u_2$, and $u_3$ are in coverage of the agent $a_4$, the agents $a_5$ and $a_6$, and the agent $a_{12}$, respectively. We assume that each user request includes the source access function, the destination access function, and two other service functions, and the user request $u_m,m=1,2,3,$ can be represented as $\{f_{m,s},f_{m,2},f_{m,3},f_{m,d}\}$. For the user request $u_1$, the neighbouring agent $a_4$ firstly updates the local network resource state and obtains the user request information. Then the agent $a_4$ makes the VNF placement decision $z_1$, where the VNFs $f_{1,2}$ and $f_{1,3}$ are deployed on the agent $a_3$ and the routing path is ${s_1} \to {a_4}(f_{1,s}) \to {a_3}({f_{1,2}},{f_{1,3}}) \to {a_4}(f_{1,d}) \to {d_1}$. For the service deployment, the agents $a_4$ and $a_3$ can share the VNF placement strategy by interacting with each other, and the agent $a_3$ will try to provide the available resources for the two VNFs $f_{1,2}$ and $f_{1,3}$ by the resource competition mechanism on the first-come-first-serve basis. If the agent $a_3$ can provide sufficient available resources for the user request $u_1$, then we can build the service chaining and provide computing service for the user request $u_1$ in a collaborative way. One of the neighbouring agents $a_5$ and $a_{10}$ makes the VNF placement decision ${z_2}$ for the user request $u_2$ as ${s_2} \to {a_{10}}(f_{2,s}) \to {a_{11}}({f_{2,2}},{f_{2,3}}) \to {a_{10}}(f_{2,d}) \to {d_2}$, and the neighbouring agent $a_{12}$ makes the VNF placement strategy decision ${z_3}$ for the user request $u_3$ as ${s_3} \to {a_{12}}(f_{3,s}) \to {a_{11}}({f_{3,2}},{f_{3,3}}) \to {a_{12}}(f_{3,d}) \to {d_3}$. However, there exists the resource conflict for the two VNF placement strategies $z_2$ and $z_3$ as the required resources for the agent $a_{11}$ are more than the available resources. The agent $a_{11}$ will provide the available resources for the user request $u_3$ that comes first, which can lead to the deployment failure of the strategy $z_2$. Then the neighbouring agent of the user request $u_2$ will update the local network resource state and make the new VNF placement decision ${z'}_2$ as ${s_2} \to {a_5}(f_{2,s}) \to {a_6}({f_{2,2}}) \to {a_2}({f_{2,3}}) \to {a_6} \to {a_5}(f_{2,d}) \to {d_2}$, where we deploy the VNF $f_{2,2}$ on the agent $a_6$ and the VNF $f_{2,3}$ on the agent $a_2$, respectively. In this instance, the user request $u_2$ can be accomplished by the three agents $a_5$, $a_6$, and $a_2$, the user request $u_3$ can be accomplished by the two agents $a_{11}$ and $a_{12}$ in a collaborative way.
\begin{algorithm}[tbp]
  \caption{Proposed Distributed VNF Placement Algorithm.}
  \label{Distributed VNF Placement Algorithm}
  \hspace*{0.02in} {\bf Input:} User requests $U$;\\
  \hspace*{0.02in} {\bf Output:} $z^{*}=\left[z_{1}^{*},z_{2}^{*},\cdots, z_{M}^{*}\right]$;

  \begin{algorithmic}[1]
  \STATE \textbf{Initialize:} $U_{next}=\varnothing, \forall u_m \in U,z_m=z_{m}^{*}=\varnothing$;
  \WHILE {$U != \varnothing$}
  \STATE All the agents share the current network resource state by a message synchronization mechanism.
  \FOR {$\forall u_m \in U$ in parallel}
  \STATE Search a feasible VNF placement strategy $z_m$ by the \emph{Viterbi} algorithm in satellite edge computing;
  \IF {$z_m == \varnothing$}
  \STATE Search a feasible VNF placement strategy $z_m$ by the \emph{Path Selection} algorithm in cloud computing;
  \ENDIF
  \ENDFOR
  \FOR {$\forall u_m \in U$ in parallel}
  \IF {$z_m != \varnothing$}
  \IF {there is no resource conflict}
  \STATE Deploy the user request $u_m$ by the strategy $z_m$ and update the network resource state;
  \STATE $z_{m}^{*} \leftarrow z_m$;
  \ELSE
  \STATE Initialize the strategy $z_m=\varnothing$;
  \STATE $U_{next} = U_{next} \cup \{u_m\}$;
  \ENDIF
  \ENDIF
  \ENDFOR
  \STATE $U \leftarrow U_{next}$ and then $U_{next}=\varnothing$;
  \ENDWHILE
  \RETURN $z^{*}$;
  \end{algorithmic}
\end{algorithm}

The proposed D-VNFP algorithm is shown in Algorithm \ref{Distributed VNF Placement Algorithm}. At the beginning, all the agents should share the current network resource state by a message synchronization mechanism. Then the source neighbouring agents for user requests will search their approximate VNF placement strategies by the \emph{Viterbi} algorithm in satellite edge computing and by the \emph{Path Selection} algorithm in cloud computing in parallel, where the \emph{Viterbi} and \emph{Path Selection} algorithms will be described in detail later. In the proposed D-VNFP algorithm, we give priority to deploying user requests on satellite edge computing by the \emph{Viterbi} algorithm for reducing the user end-to-end delay. If the available resources on satellite edge servers are insufficient, then we will consider to deploy user requests to cloud servers by the \emph{Path Selection} algorithm. For the service deployment, the VNF placement strategies for user requests can be executed on the agents simultaneously. When there exists resource conflict for the user request $u_m$, we will initialize the strategy as $z_m=\varnothing$ and put the user request $u_m$ into the list $U_{next}$, which is initialized as $U_{next}=\varnothing$. When there is no resource conflict for the user request $u_m$, then we can deploy the user request $u_m$ to satellite edge and cloud by the strategy $z_{m}^{*}=z_m$ and update the network state. After all user requests are deployed, we will reconsider the set of user requests as $U \leftarrow U_{next}$ and perform the proposed D-VNFP algorithm again. When none of user requests needs to be deployed in satellite edge and cloud computing, that is, $U== \varnothing$, the iteration will terminate. When edge and cloud servers have insufficient resources for deploying user requests, these user requests will be performed locally.

For the VNF placement problem in satellite edge computing, the \emph{Viterbi} algorithm \cite{7469866,Potentialgame_sat_IoT}, is used to jointly minimize the satellite network bandwidth usage and the user end-to-end delay, where the \emph{Viterbi} algorithm is a heuristic optimization algorithm and can find a near-optimal solution in a relatively shorter time. In order to optimize the network bandwidth cost while guaranteeing the minimum user end-to-end delay, one of the source neighbouring agents for a user request firstly obtains the $d$ shortest paths between the source and the destination. To be more specific, one source neighbouring agent of the user request $u_m$ can be randomly considered as the decision-making node to calculate the path end-to-end delay by traversing the source and destination neighbouring agents. The total number of traversed paths for the user request $u_m$ is indicated as $N(u_m,s)\cdot N(u_m,d) \cdot d$ and the $d$ shortest paths will remain to be the candidate paths for further processing. We assume that the number of paths remaining for the source neighbouring agent $a_n$ of the user request $u_m$ is represented as $P_{s_m,a_n}^{d_m}$, then $\bigcup\nolimits_{{a_n}} {P_{{s_m},{a_n}}^{{d_n}}}  = P_{{s_n}}^{{d_n}},{a_n} \in N({u_m},s)$. We sort the remaining paths $P_{s_m}^{d_m}$ by delay cost in ascending order. The source neighbouring agent can traverse the remaining paths from low delay to high delay and then deploy the VNFs by the \emph{Viterbi} algorithm to optimize the network bandwidth cost.

For the \emph{Viterbi} algorithm, each deployment strategy of the VNF is considered as a possible state $\varphi$ and all possible states for the same VNF make up the state set $\Omega$ of the VNF in a stage. The edge between different states from adjacent stages represents the network bandwidth cost, and the cumulative bandwidth cost $c_{bw}^{m}$ for the VNF placement strategy is saved in the state node. The number of stages for a user request is equal to the number of the VNFs. After all the VNFs are deployed, the final cumulative network bandwidth costs can be found in the last destination state nodes, where the most likely path with the minimum bandwidth cost can be considered as the optimal VNF placement strategy. Note that the \emph{Viterbi} algorithm is performed under the multiple physical constraints as shown in the equations \eqref{equation1}-\eqref{equation24}. Furthermore, To reduce the computational complexity of the \emph{Viterbi} algorithm, we remain the $B$ optimal possible states of each stage to participate in the VNF placement decisions of the next stage. When the source neighbouring agent finds the first VNF placement strategy $z_m$ with the minimum bandwidth and delay costs, the search procedure for the user request $u_m$ will terminate. The procedure of the \emph{Viterbi} algorithm is shown in Algorithm \ref{Viterbi Algorithm}.
\begin{algorithm}[tbp]
  \caption{Viterbi Algorithm.}
  \label{Viterbi Algorithm}
  \hspace*{0.02in} {\bf Input:} $u_m$;\\
  \hspace*{0.02in} {\bf Output:} $z_m$;
  \begin{algorithmic}[1]
  \STATE \textbf{Initialize:} $z_m=\varnothing,\forall a_n \in N(u_m,s), z_m(n)=\varnothing$;
  \STATE Obtain the paths $P_{s_m,a_n}^{d_m},a_n \in N(u_m,s)$;
  \STATE Sort all the paths by end-to-end delay in ascending order;
  \FOR {each $p \in P_{s_m}^{d_m}$}
  \IF {the delay constraint is not satisfied}
  \STATE Continue;
  \ENDIF
  \STATE Sort the VNFs except $f_{m,s}$ by topology as $\Gamma_{m}$;
  \STATE $\Omega=\{\varphi_{m,s}\}$;
  \FOR {each $f_{m,i} \in \Gamma_{m}$}
  \STATE $\Omega'=\varnothing$;
  \FOR {each $\varphi \in \Omega$}
  \STATE Update the current network resource state by $\varphi$;
  \STATE List the candidate edge servers $V_{m,i}^{p}$;
  \FOR {each $v_n \in V_{m,i}^{p}$}
  \STATE Calculate the required resources;
  \IF {the required resources are satisfied}
  \STATE Obtain the bandwidth cost $c_{bw}^{m}$ and the VNF deployment strategy $\varphi'$;
  \STATE $\Omega' = \Omega' \cup \{\varphi,\varphi'\}$;
  \ENDIF
  \ENDFOR
  \ENDFOR
  \STATE Sort $\Omega'$ by bandwidth cost in ascending order;
  \STATE $\Omega \leftarrow \Omega'[:B]$;
  \ENDFOR
  \IF {$\Omega != \varnothing$}
  \STATE Obtain the optimal strategy $z_m$ and break;
  \ENDIF
  \ENDFOR
  \RETURN $z_m$;
  \end{algorithmic}
\end{algorithm}

For the VNF placement problem in remote cloud computing, we assume that all the VNFs except the source access function and the destination access function for a user request will be deployed to cloud servers and the processing delay of switches and links in the cloud data center is not considered. The satellite network is just responsible for routing original data from the source to the cloud data center and returning the processing results from the cloud data center to the destination. In this scenario, our aim is to find the best available paths between the source and the cloud data center and the cloud data center and the destination, respectively. Then we will route the data flow produced by a user request to the cloud data center for further processing and the processing results will be sent back to the destination by the satellite network.
\begin{algorithm}[tbp]
  \caption{Path Selection Algorithm.}
  \label{Path Selection Algorithm}
  \hspace*{0.02in} {\bf Input:} $u_m$;\\
  \hspace*{0.02in} {\bf Output:} $z_m$;
  \begin{algorithmic}[1]
  \STATE \textbf{Initialize:} $z_{s_m}^{dc}=z_{dc}^{d_m}=\varnothing$;
  \STATE Obtain $P_{s_m,a_n}^{dc},a_n \in N(u_m,s),P_{dc,a_n}^{d_m},a_n \in N(dc)$;
  \STATE Sort all the paths by end-to-end delay in ascending order;
  \FOR {each $p \in P_{s_m}^{dc}$}
  \STATE Calculate the bandwidth and service requirements;
  \IF {the required constraints are satisfied}
  \STATE Obtain the optimal path strategy $z_{s_m}^{dc}$ and break;
  \ENDIF
  \ENDFOR
  \FOR {each $p \in P_{dc}^{d_m}$}
  \STATE Calculate the bandwidth and service requirements;
  \IF {the required constraints are satisfied}
  \STATE Obtain the optimal path strategy $z_{dc}^{d_m}$ and break;
  \ENDIF
  \ENDFOR
  \IF {$z_{s_m}^{dc}!= \varnothing$ and $z_{dc}^{d_m}!= \varnothing$}
  \STATE $z_m=\{z_{s_m}^{dc},z_{dc}^{d_m}\}$;
  \ENDIF
  \RETURN $z_m$;
  \end{algorithmic}
\end{algorithm}

In this paper, we use the \emph{Path Selection} algorithm to find the optimal available path with the minimum end-to-end delay cost for a user request. The procedure of the \emph{Path Selection} algorithm is shown in Algorithm \ref{Path Selection Algorithm}. For the user request $u_m$, initially, we consider the neighbouring agents $N(u_m,s)$ of the source as the access points and calculate the path end-to-end delay from the source to the cloud data center by traversing the neighbouring agents of the cloud data center, respectively. Then we can obtain $N(u_m,s)\cdot N(dc)\cdot d$ paths between the source and the cloud data center and sort all the paths by end-to-end delay in ascending order to select the $d$ shortest paths. We indicate the remaining paths for the source neighbouring agent $a_n$ as $P_{s_m,a_n}^{dc}$ and then $\bigcup\nolimits_{{a_n}} {P_{{s_m},{a_n}}^{{dc}}}  = P_{{s_n}}^{{dc}},{a_n} \in N({u_m},s)$. Similarly, we can obtain the $d$ shortest paths between the cloud data center and the destination and indicate the remaining paths for the neighbouring agent $a_n$ of the cloud data center as $P_{dc,a_n}^{d_m}$, where $\bigcup\nolimits_{{a_n}} {P_{{dc},{a_n}}^{{d_m}}}  = P_{{dc}}^{{d_m}},{a_n} \in N(dc)$. The source neighbouring agent can traverse the paths $P_{s_m}^{dc}$ from low delay to high delay and calculate the network bandwidth and service requirements. If the bandwidth and service requirements are satisfied, we can consider the current path as the optimal path strategy $z_{s_m}^{dc}$ and break the search procedure. For the neighbouring agents of the cloud data center, similar to the procedure of finding the optimal path strategy for the source neighbouring agents, we can obtain the optimal path strategy $z_{dc}^{d_m}$. If $z_{s_m}^{dc}!=\varnothing$ and $z_{dc}^{d_m}!=\varnothing$, the optimal VNF placement strategy can be indicated as $z_m=\{z_{s_m}^{dc},z_{dc}^{d_m}\}$.

By using the proposed D-VNFP algorithm to address the VNF placement problem in equations \eqref{equation27} and \eqref{equation28}, we improve the network bandwidth cost and the user end-to-end delay compared with the Greedy and Viterbi algorithms.

\subsection{Algorithm Complexity}\label{Algorithm Complexity}
For the proposed D-VNFP algorithm, we assume that $M$ user requests need to be provided computing services in a time slot. In the worst case, only one user request is successfully deployed to satellite edge and cloud in each iteration because of resource competition. Thus, the procedure of VNF placement for user requests should be run for $\frac{M(M+1)}{2}$ times. The computation complexity of the \emph{Viterbi} and \emph{Path selection} algorithms can be indicated as $O(dBNF)$ and $O(d)$, where we indicate the maximum number of VNFs for a user request as $F$. Therefore, we can obtain the computation complexity of the proposed D-VNFP algorithm to be $O(M^2dBNF)$. However, the computation complexity of the two baseline algorithms of Viterbi and Greedy can be indicated as $O(MdBNF)$ and $O(MdNF)$. The main difference between the proposed D-VNFP algorithm and the two baseline algorithms is that the proposed D-VNFP algorithm is a distributed algorithm and without a centralized controller, but the two baseline algorithms are centralized algorithms and a centralized controller is necessary to obtain the network state and the information about user requests.

\section{Performance Evaluation}\label{Performance Evaluation}
For the VNF placement problem in satellite edge and cloud computing, we make the experiments to evaluate the performance of the proposed D-VNFP algorithm in static and dynamic environments, respectively. Two existing centralized optimization algorithms, i.e., Greedy \cite{7332796} and Viterbi \cite{7469866}, are considered as the baseline algorithms, which are used to address the VNF placement problem. For Greedy, we traverse the $d$ shortest paths between the source and the destination by end-to-end delay in ascending order and also deploy the VNFs for a user request by a greedy way. The Viterbi-only procedure, as shown in Algorithm \ref{Viterbi Algorithm}, is considered as the other baseline algorithm. As shown in the simulation results, the proposed D-VNFP algorithm outperforms the two baseline algorithms of Greedy and Viterbi.

\subsection{Simulation Setup}
\begin{table}[tbp]
  \centering
  \caption{Parameter Setting}
  \label{Parameter Setting}
  \resizebox{\columnwidth}{!}{
    \begin{tabular}{p{0.45\columnwidth}<{\centering}p{0.45\columnwidth}<{\centering}}
    \hline
    \multicolumn{2}{c}{Satellite Network\cite{8975779}} \\
    \hline
    Number of satellites & 12 \\
    Inter-satellite link bandwidths & 1 Gbps \\
    Inter-satellite link delay & 7.25 ms, 12.6 ms, 13.4 ms \\
    Satellite-ground link bandwidths & 10 Gbps \\
    Satellite-ground link delay & 13.1 ms \\
    Edge server & 96 vCPUs, 112 GB memory \\
    \hline
    \multicolumn{2}{c}{User Requests} \\
    \hline
    Number of VNFs & Truncated power-law distribution \\
    Required resources per VNF & [1,2] vCPUs, [2,4] GB memory \\
    Required bandwidths per edge & [1,5] Mbps \\
    Computing time per VNF & [20,30] ms \\
    Running period per user request & Exponent distribution \\
    Number of user requests & Poisson distribution \\
    \hline
    \end{tabular}%
    }
\end{table}%
\begin{figure}[tbp]
  \centering
  \subfigure[Network bandwidth cost]{\includegraphics[width=0.35\textwidth]{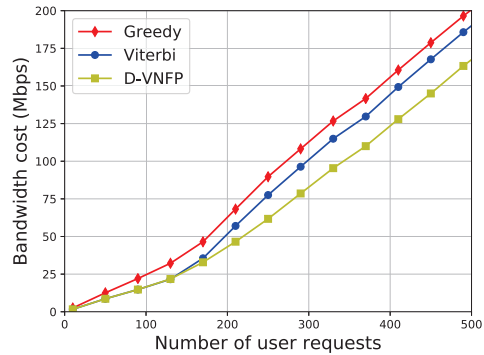}
  \label{Network bandwidth cost in static environment}}
  \subfigure[User end-to-end delay cost]{\includegraphics[width=0.35\textwidth]{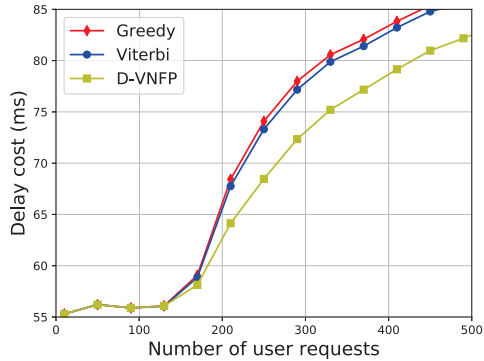}
  \label{User end-to-end delay cost in static environment}}
  \caption{Performance comparison for D-VNFP, Greedy, and Viterbi.}
  \label{Performance comparison for D-VNFP, Greedy, and Viterbi in static environment}
\end{figure}
A satellite network with $N=12$ LEO satellites is used in our simulation experiments, where there are 3 orbital planes and each orbital plane has 4 LEO satellites. For each satellite, the altitude is 780 km, the available resources include 96 vCPUs and 112 GB memory, and the number of inter-satellite links is 4. We assume that the delay time of the inter-satellite link for the same orbital plane is 7.25 ms and 12.6 ms, respectively, for the adjacent orbital planes is 13.4 ms, and for the satellite-ground link is 13.1 ms \cite{8975779}. The bandwidth capacity of each inter-satellite link is 1 Gbps. We assume that the cloud data center is in the coverage of the satellites $v_6$ and $v_7$, where there are sufficient CPU and memory resources and the satellite-ground link bandwidth capacity is 10 Gbps.

For a user request, we assume the number of the VNFs follows a truncated power-law distribution, where the exponent is 2, the minimum value is 2 and the maximum value is 7, respectively \cite{RankothgeLRL17}. The resource requirements for each VNF are randomly generated from $[1,2]$ vCPUs and $[2,4]$ GB memory, and the computing time for each VNF is $[20,30]$ ms. The bandwidth requirements for each edge between the neighbouring VNFs are $[1,5]$ Mbps. The source and the destination are randomly generated in the coverage of satellites. We also assume that the maximum acceptable delay time for each user request can guarantee that user requests can be offloaded to the cloud data center for further processing. In dynamic environment, the running time of each user request follows the Exponent distribution with $\lambda_E=3$ and the number of user requests per time slot follows the Poisson distribution. The main simulation parameters are summarized in Table \ref{Parameter Setting}. According to the parameter evaluation results in \cite{Potentialgame_sat_IoT}, we set the number of the shortest paths between two satellites as $d=8$ and the search width of the Viterbi algorithm as $B=4$.
\begin{figure*}[tbp]
  \centering
  \subfigure[Network bandwidth cost]{\includegraphics[width=0.3\textwidth]{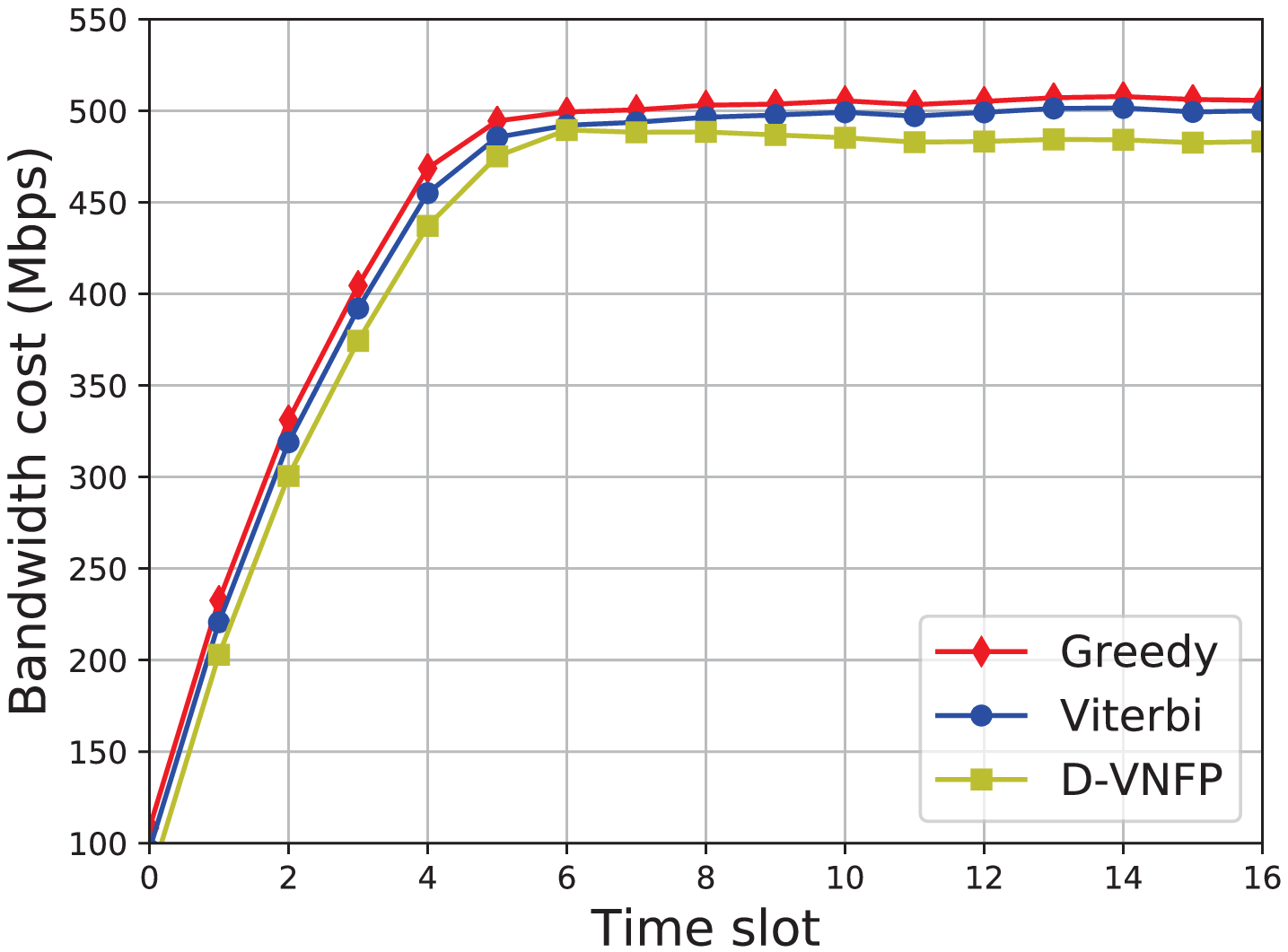}
  \label{Network bandwidth cost for 290 user requests}}
  \subfigure[User end-to-end delay cost]{\includegraphics[width=0.3\textwidth]{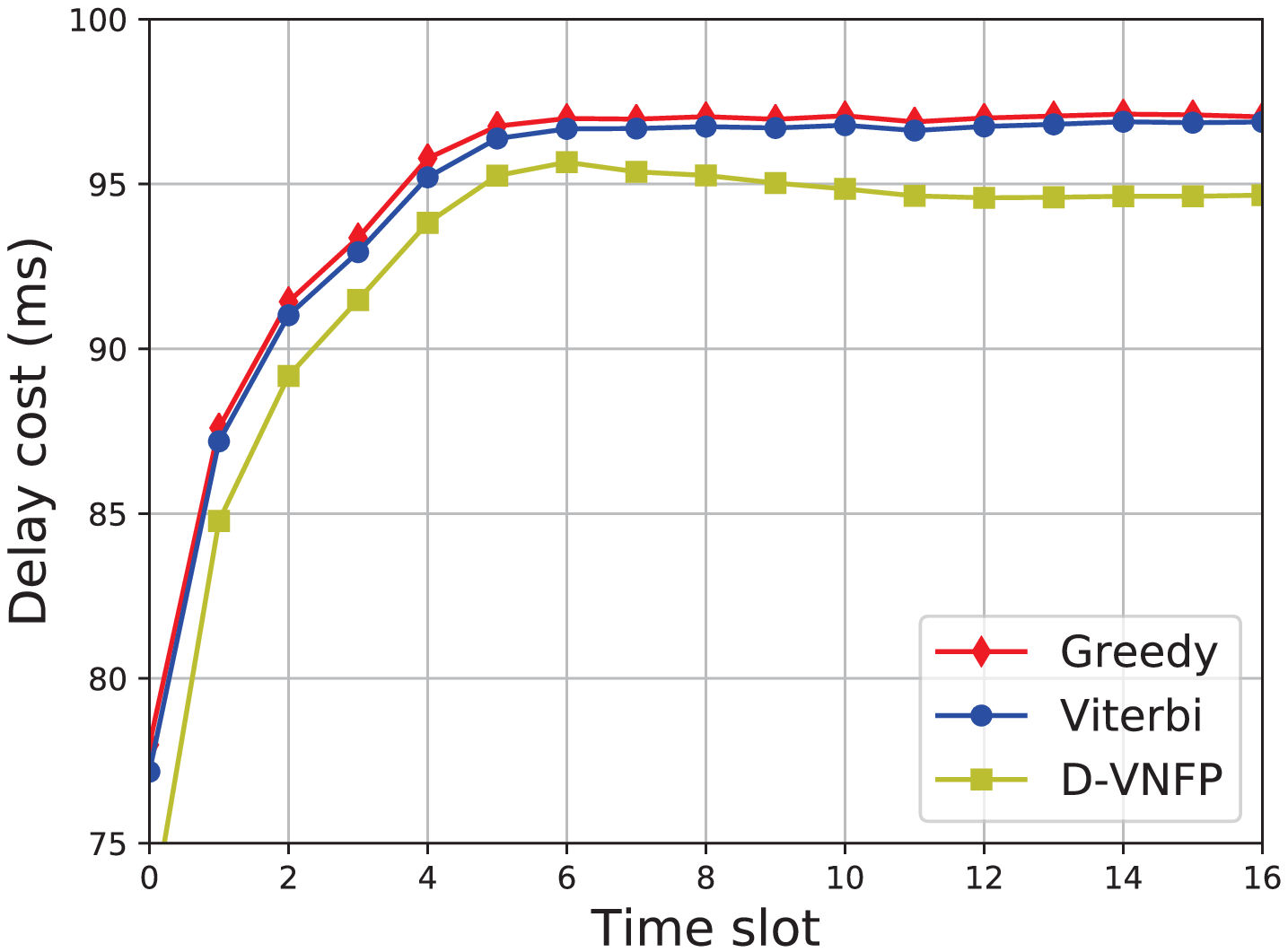}
  \label{User end-to-end delay cost for 290 user requests}}
  \subfigure[Percentage of allocated user requests]{\includegraphics[width=0.3\textwidth]{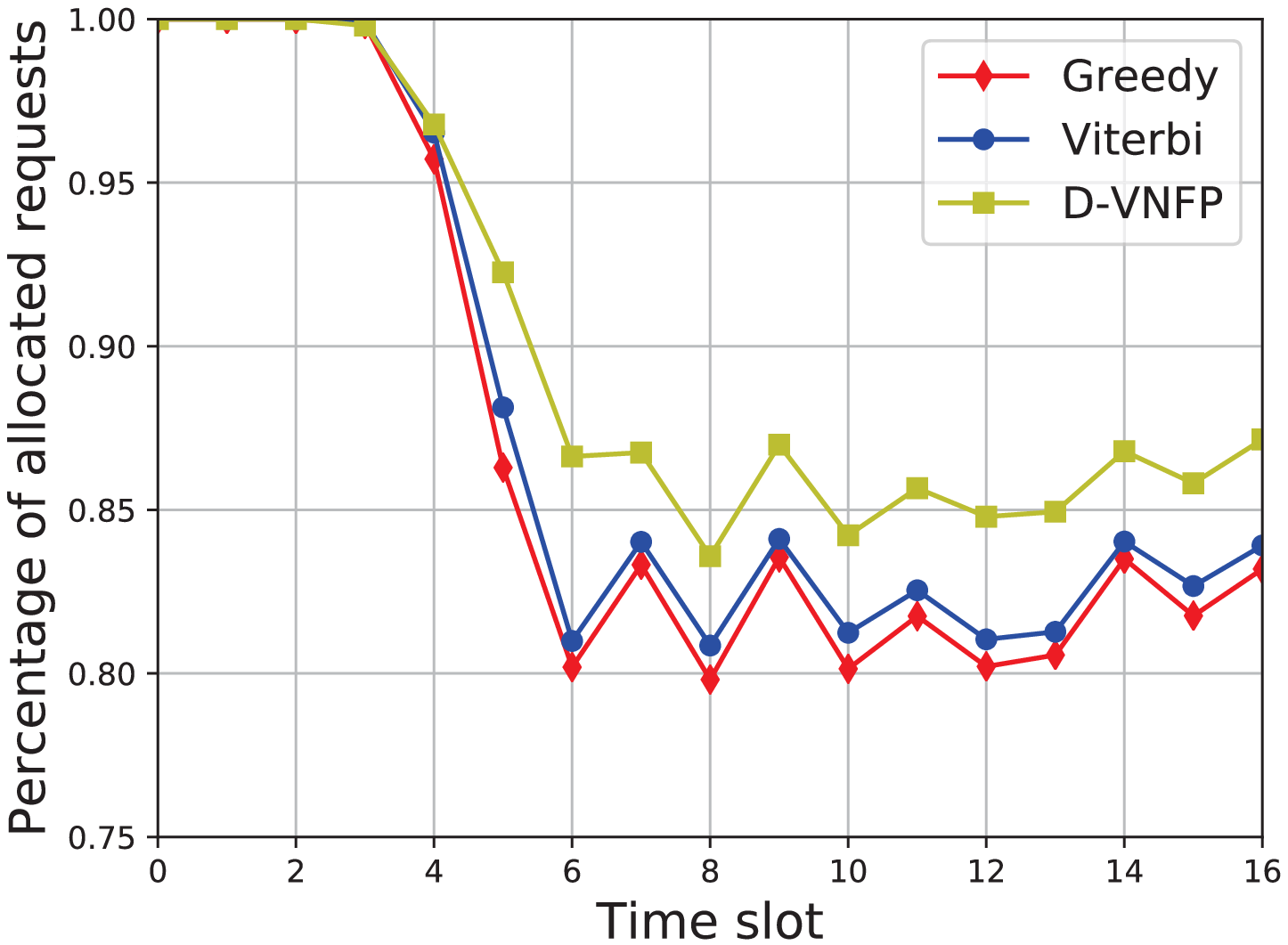}
  \label{Percentage of allocated user requests for 290 user requests}}
  \caption{Simulation results for $\lambda_P = 290$ in dynamic environment.}
  \label{Simulation results for 290 user requests in dynamic environment}
\end{figure*}

\subsection{Performance Comparison with the Baseline Algorithms}
Based on the above simulation parameters, we conduct the following experiments to evaluate the performance of the proposed D-VNFP algorithm by comparing with the two baseline algorithms of Greedy and Viterbi. The number of user requests for the experiments is set as $M=\{10,30,\cdots,590\}$, where all user requests for each experiment can be deployed to the satellite edge and cloud. Each experiment is run for 30 times and we obtain the average results in terms of network bandwidth and user end-to-end delay.

The simulation results for different number of user requests are shown in Fig.~\ref{Performance comparison for D-VNFP, Greedy, and Viterbi in static environment}. Fig.~\ref{Network bandwidth cost in static environment} illustrates the average network bandwidth costs for different user requests. We can observe that the proposed D-VNFP algorithm outperforms the Greedy algorithm for all cases. When the number of user requests is small, e.g., $M=70$, 90, and 110, the proposed D-VNFP algorithm performs similar to the Viterbi algorithm. As the number of user requests increases, the performance of the proposed D-VNFP algorithm is better than that of the Viterbi algorithm. In the case of $M=110$, the bandwidth costs for D-VNFP, Greedy, and Viterbi are 19.10 Mbps, 27.99 Mbps, and 18.98 Mbps, respectively, the average bandwidth cost obtained by the proposed D-VNFP algorithm reduces by $31.75\%$ for Greedy. For $M=290$, the average bandwidth cost obtained by the proposed D-VNFP algorithm reduces by $27.38\%$ for Greedy and $18.43\%$ for Viterbi, respectively. For all cases, the proposed D-VNFP algorithm can reduce the average bandwidth cost by $21.14\%$ for Greedy and $13.51\%$ for Viterbi, respectively.

Fig.~\ref{User end-to-end delay cost in static environment} describes the user end-to-end delay costs for different user requests. We can observe that the performance of the proposed D-VNFP algorithm is very close to that of Greedy and Viterbi for the small number of user requests, e.g., $M=70$, 90, and 110. As the number of user requests increases, the proposed D-VNFP algorithm outperforms Greedy and Viterbi, where the performance of the Viterbi algorithm is slightly better than that of the Greedy algorithm. That is due to the fact that the satellites can provide sufficient available computing resources for user requests with small $M$, but the available resources of satellites will be less than the resource requirements of user requests as $M$ increases and then the user requests will be deployed to the remote cloud data center, which can lead to the high network transmission delay. For $M=290$, the user end-to-end delay costs for D-VNFP, Greedy, and Viterbi are 72.34 ms, 77.98 ms, and 77.16 ms, respectively. The user end-to-end delay obtained by the proposed D-VNFP algorithm reduces by $7.23\%$ for Greedy and $6.24\%$ for Viterbi. On average, the proposed D-VNFP algorithm can improve the performance of the user end-to-end delay by $4.38\%$ for Greedy and $3.80\%$ for Viterbi, respectively.

\subsection{VNF placement in Dynamic Environment}
To evaluate the on-line performance of the proposed D-VNFP algorithm for deploying user requests, we conduct the following experiments in dynamic environment. We assume that the number of user requests to come in a time slot follows the Poisson distribution with $\lambda_P=\{10,30,\cdots,310\}$ and the number of time slots for each experiment is 50. We run each experiment for 30 times and obtain the average results.

For an example of $\lambda_P = 290$, the simulation results for different time slots are shown in Fig.~\ref{Simulation results for 290 user requests in dynamic environment}. Fig.~\ref{Network bandwidth cost for 290 user requests} depicts the network bandwidth costs for deploying the user requests over different time slots. We can observe that the proposed D-VNFP algorithm performs better than the two baseline algorithms of Greedy and Viterbi for all time slots. The performance improvement of the proposed D-VNFP algorithm for Greedy and Vertbi is $4.78\%$ and $3.26\%$, respectively. Fig.~\ref{User end-to-end delay cost for 290 user requests} describes the user end-to-end delay costs for deploying the user requests over different time slots. We can observe that the performance of the proposed D-VNFP algorithm is better than that of Greedy and Viterbi. The user end-to-end delay obtained by the proposed D-VNFP algorithm reduces by $2.54\%$ for Greedy and $2.24\%$ for Viterbi, respectively. Fig.~\ref{Percentage of allocated user requests for 290 user requests} illustrates the percentages of allocated user requests over different time slots. As the number of user requests deployed to the satellite edge and cloud increases, the satellite network can not meet the resource requirements of new user requests, e.g., CPU, memory, and bandwidth. For this reason, some user requests may fail to deploy to the satellite edge and cloud and the percentages of allocated user requests will decrease. However, the proposed D-VNFP algorithm also outperforms the two baseline algorithms of Greedy and Viterbi. The performance improvement of the proposed D-VNFP algorithm is $4.19\%$ for Greedy and $3.29\%$ for Viterbi, respectively.
\begin{figure*}[tbp]
  \centering
  \subfigure[Network bandwidth cost]{\includegraphics[width=0.3\textwidth]{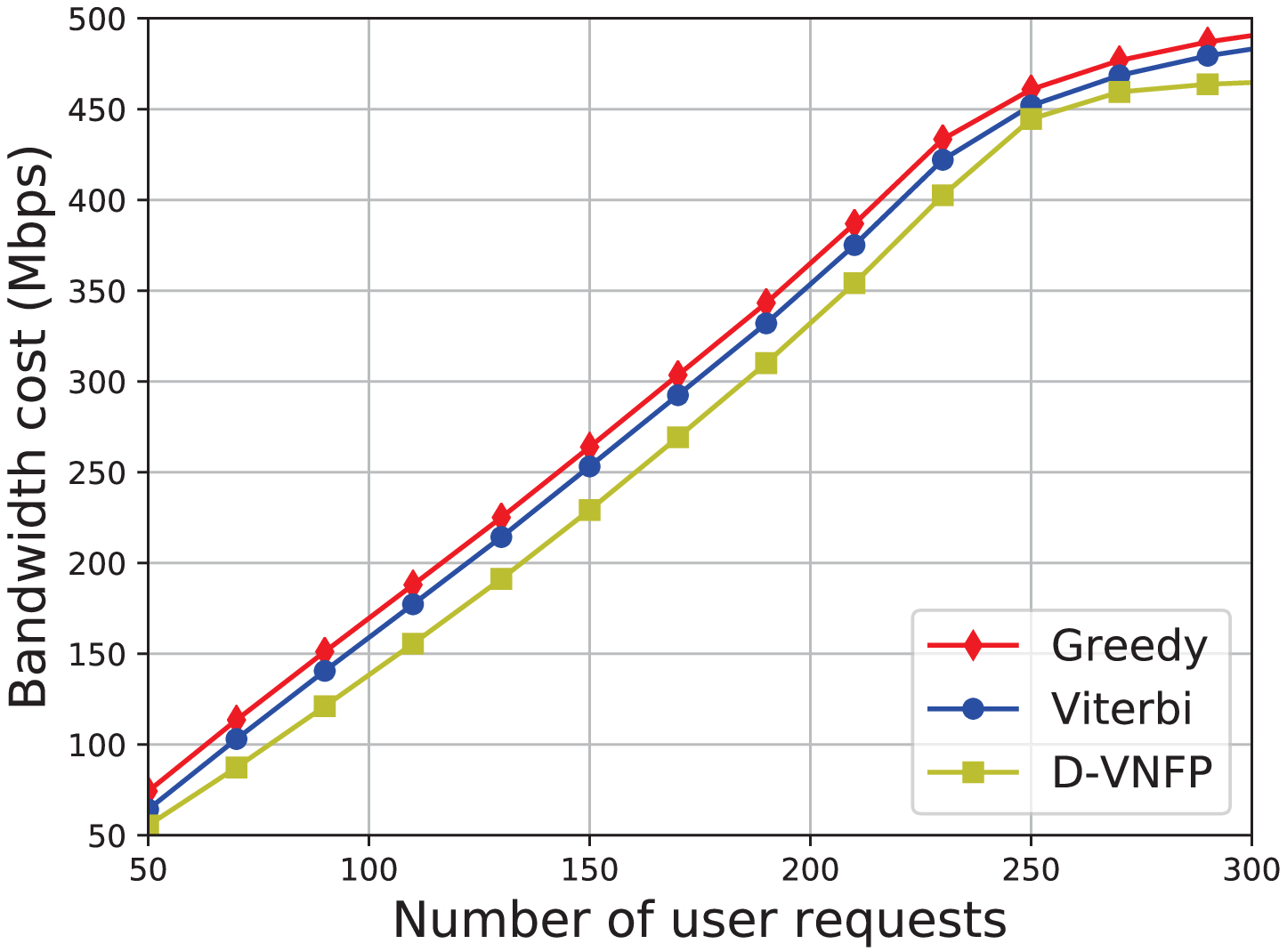}
  \label{Network bandwidth cost in dynamic environment}}
  \subfigure[User end-to-end delay cost]{\includegraphics[width=0.3\textwidth]{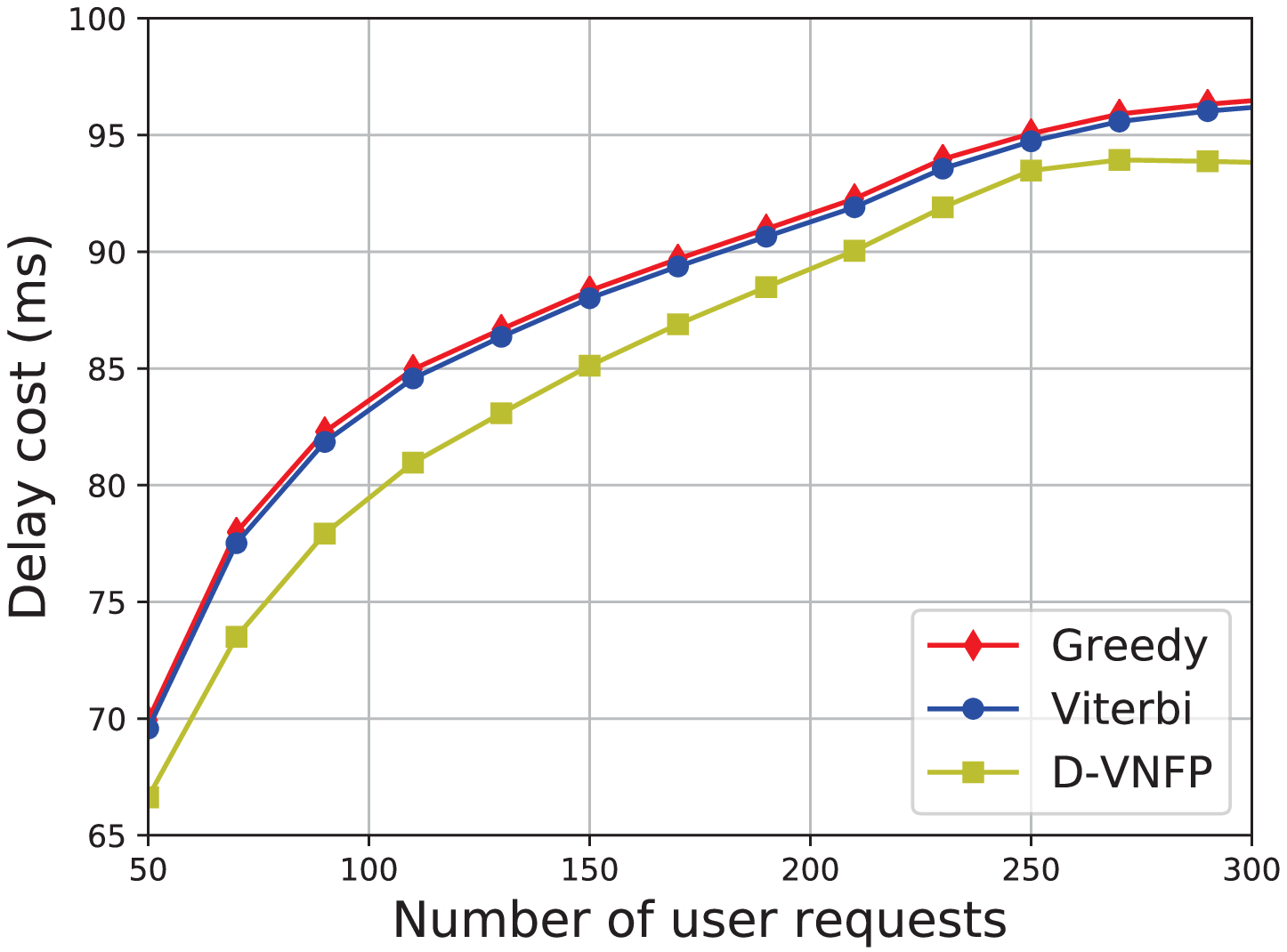}
  \label{User end-to-end delay cost in dynamic environment}}
  \subfigure[Percentage of allocated user requests]{\includegraphics[width=0.3\textwidth]{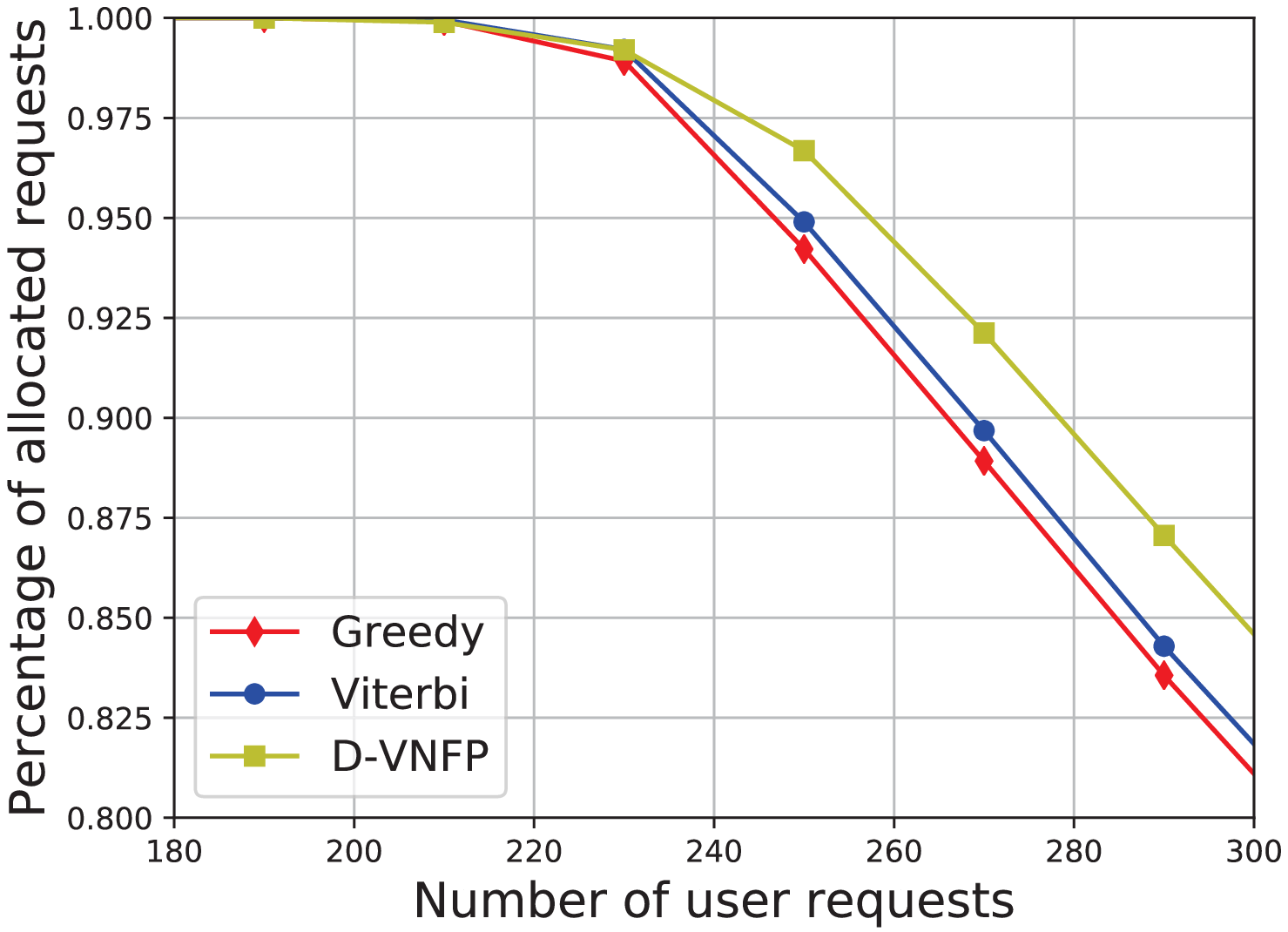}
  \label{Percentage of allocated user requests in dynamic environment}}
  \caption{Simulation results for different number of user requests in dynamic environment.}
  \label{Simulation results for different number of user requests in dynamic environment}
\end{figure*}

In addition, to further demonstrate the effectiveness of the proposed D-VNFP algorithm, we provide the average simulation results for different user requests in dynamic environment, as shown in Fig.~\ref{Simulation results for different number of user requests in dynamic environment}. Fig.~\ref{Network bandwidth cost in dynamic environment} describes the average network bandwidth costs for different user requests in dynamic environment. We can observe that the proposed D-VNFP algorithm outperforms the two baseline algorithms of Greedy and Viterbi for all cases, where the Greedy algorithm has the worst performance and is followed by the Viterbi algorithm. In the case of $\lambda_P=90$, the network bandwidth costs for D-VNFP, Greedy, and Viterbi are 121.05 Mbps, 151.11 Mbps, and 140.54 Mbps, respectively. The performance improvement of the proposed D-VNFP algorithm is $19.89\%$ for Greedy and $13.86\%$ for Viterbi. On average, the network bandwidth cost obtained by the proposed D-VNFP algorithm reduces by $9.12\%$  for Greedy and $5.87\%$ for Viterbi, respectively. Fig.~\ref{User end-to-end delay cost in dynamic environment} illustrates the average user end-to-end delay costs for different user requests in dynamic environment. It is observed that the proposed D-VNFP algorithm performs better than the two baseline algorithms of Greedy and Viterbi. For an example of $\lambda_P=90$, the user end-to-end delay costs for D-VNFP, Greedy, and Viterbi are 77.92 ms, 82.28 ms, and 81.85 ms, respectively. The user end-to-end delay obtained by the proposed D-VNFP algorithm reduces by $5.30\%$  for Greedy and $4.79\%$ for Viterbi. For all cases, The proposed D-VNFP algorithm reduces the user end-to-end delay by $3.06\%$ for Greedy and $2.70\%$ for Viterbi on average.

Fig.~\ref{Percentage of allocated user requests in dynamic environment} describes the average percentages of allocated user requests in dynamic environment. We can observe that the satellite edge and cloud can provide available computing resources for user requests when the number of user requests is small, and then all user requests can be deployed to the satellite edge and cloud for the proposed D-VNFP algorithm and the two baseline algorithms. As the number of user requests increases, the available computing and bandwidth resources of the satellite network are gradually decreasing. When the required resources of user requests can not be satisfied by the satellite network, the user requests can not be deployed to the satellite edge and cloud. Due to the fact that the performance of the Greedy algorithm is the worst, the percentage of allocated user requests obtained by the Greedy algorithm firstly begins to decline and is followed by the Viterbi algorithm. For an example of $\lambda_P=290$, the percentages of allocated user requests for D-VNFP, Greedy, and Viterbi are 0.87, 0.83, and 0.84, respectively. The performance improvement of the proposed D-VNFP algorithm is $4.19\%$ for Greedy and $3.29\%$ for Viterbi. For all cases, the average percentage of allocated user requests of the proposed D-VNFP algorithm improves by $0.83\%$  for Greedy and $0.62\%$ for Viterbi, respectively.

\section{Conclusion}\label{Conclusion}
In this paper, we investigate the VNF placement problem in collaborative satellite edge and cloud computing with jointly minimizing the satellite network bandwidth usage and the user end-to-end delay, and we formulate the VNF placement problem as an integer non-linear programming problem. As the VNF placement problem is NP-hard, we propose a distributed VNF placement algorithm to address the problem. For each user request, we use the \emph{Viterbi} algorithm and the \emph{Path Selection} algorithm to find an approximate VNF placement solution in satellite edge and cloud computing, respectively. All user requests can be deployed to satellite edge and cloud in a distributed manner and we use a resource competition mechanism based on the first-come-first-serve basis to solve the possible resource conflict problem.

To evaluate the performance of the proposed D-VNFP algorithm, we conduct the experiments in static and dynamic environments and compare the performance with two baseline centralized algorithms of Greedy and Viterbi. The simulation results show the proposed D-VNFP algorithm performs better than the two baseline algorithms. Specifically, in static environment, the proposed D-VNFP algorithm improves the network bandwidth cost by $21.14\%$ for Greedy and $13.51\%$ for Viterbi, and reduces the user end-to-end delay cost by $4.38\%$ for Greedy and $3.80\%$ for Viterbi, respectively. In dynamic environment, the proposed D-VNFP algorithm reduces the network bandwidth cost by $9.12\%$  for Greedy and $5.87\%$ for Viterbi, reduces the user end-to-end delay cost by $3.06\%$ for Greedy and $2.70\%$ for Viterbi, and improves the average percentage of allocated user requests by $0.83\%$ for Greedy and $0.62\%$ for Viterbi, respectively.


%

%

%
%

\ifCLASSOPTIONcaptionsoff
  \newpage
\fi



\bibliographystyle{IEEEtran}
\bibliography{A_Distributed_Virtual_Network_Function_Placement_Approach_in_Satellite_Edge_and_Cloud_Computing}
%
%
%

%



\begin{IEEEbiography}[{\includegraphics[width=1in,height=1.25in,clip,keepaspectratio]{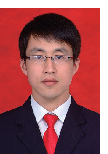}}]{Xiangqiang Gao} received the B.Sc. degree in the School of Electronic Engineering from Xidian University and the M.Sc. degree from Xi\textquoteright an Microelectrinics Technology Institute, Xi\textquoteright an, China, in 2012 and 2015, respectively. He is currently pursuing the Ph.D. degree with the School of Electronic and Information Engineering, Beihang University, Beijing, China. His research interests include rateless codes, software defined network and network function virtualization.\par
\end{IEEEbiography}

\begin{IEEEbiography}[{\includegraphics[width=1in,height=1.25in,clip,keepaspectratio]{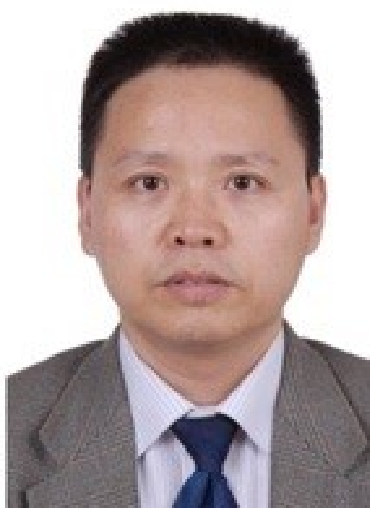}}]{Rongke Liu} received the B.S. and Ph.D. degrees from Beihang University in 1996 and 2002, respectively. He was a Visiting Professor with the Florida Institution of Technology, USA, in 2006; The University of Tokyo, Japan, in 2015; and the University of Edinburgh, U.K., in 2018, respectively. He is currently a Full Professor with the School of Electronic and Information Engineering, Beihang University. He received the support of the New Century Excellent Talents Program from the Minister of Education, China. He has attended many special programs, such as China Terrestrial Digital Broadcast Standard. He has published over 100 papers in international conferences and journals. He has been granted over 20 patents. His research interest covers wireless communication and space information network.\par
\end{IEEEbiography}
\begin{IEEEbiography}[{\includegraphics[width=1in,height=1.25in,clip,keepaspectratio]{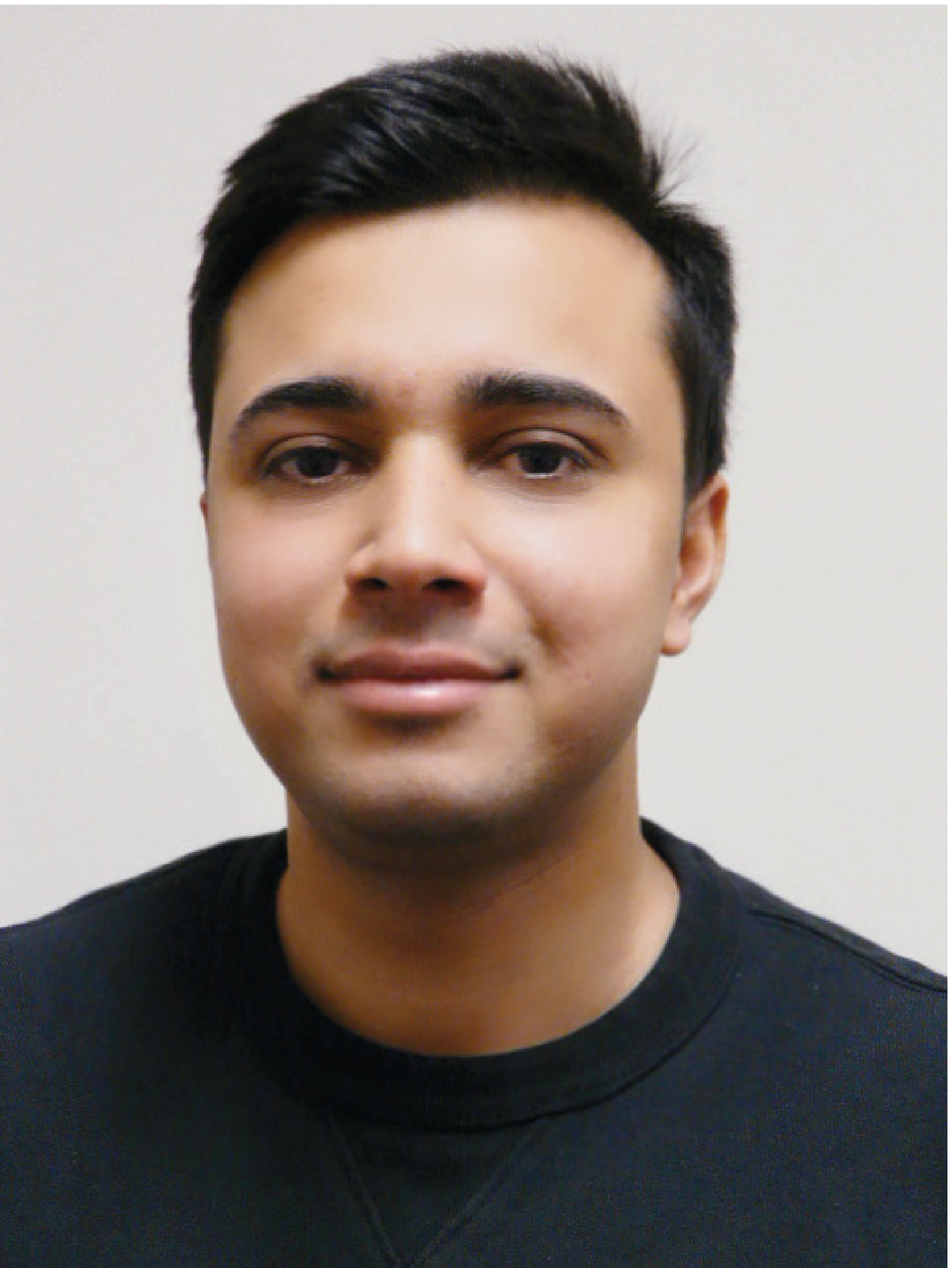}}]{Aryan Kaushik} is currently a Research Fellow at the Department of Electronic and Electrical Engineering, University College London (UCL), United Kingdom, from Feb. 2020. He received PhD in Communications Engineering at the Institute for Digital Communications, School of Engineering, The University of Edinburgh, United Kingdom, in Jan. 2020. He received MSc in Telecommunications from The Hong Kong University of Science and Technology, Hong Kong, in 2015. He has held visiting research appointments at the Wireless Communications and Signal Processing Lab, Imperial College London, UK, from 2019-20, the Interdisciplinary Centre for Security, Reliability and Trust, University of Luxembourg, Luxembourg, in 2018, and the School of Electronic and Information Engineering, Beihang University, China, from 2017-19. His research interests are broadly in signal processing, radar, wireless communications, millimeter wave and multi-antenna communications.\par
\end{IEEEbiography}

\begin{IEEEbiography}[{\includegraphics[width=1in,height=1.25in,clip,keepaspectratio]{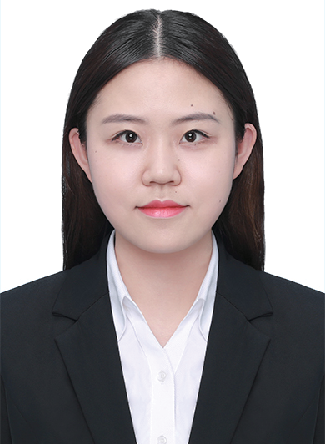}}]{Hangyu Zhang} received the B.S. degree in the School of Communication Engineering from Jilin University in 2019. She has been working on the Ph.D. degree in the School of Electronic and Information Engineering, Beihang University, Beijing, China. Her research interest includes the development of machine learning in Satellite Internet.\par
\end{IEEEbiography}

\vfill


\end{document}